\documentclass{aa}

\usepackage[varg]{txfonts}
\usepackage{graphicx}
\usepackage{algorithm}
\usepackage{algorithmic}
\usepackage{color}
\usepackage{placeins}
\usepackage{subcaption}
\usepackage{CJK}
\usepackage{amsmath}
\usepackage{natbib,twoopt}
\usepackage[breaklinks=true]{hyperref}
\usepackage{comment}
\usepackage{adjustbox}
\usepackage{color}
\usepackage{stfloats}
\bibpunct{(}{)}{;}{a}{}{,}

\authorrunning{Ruan et al.}

\begin{document}
\begin{CJK*}{UTF8}{gbsn}

\title{Investigation on deep learning-based galaxy image translation models}

\author{Hengxin Ruan (阮恒心)\inst{\ref{inst1}}\thanks{E-mail: ruanhx@pcl.ac.cn}
\and Qiufan Lin (林秋帆)\inst{\ref{inst1}}\thanks{Corresponding author; E-mail: linqf@pcl.ac.cn}
\and Shupei Chen (陈树沛)\inst{\ref{inst1}}
\and Yang Wang (汪洋)\inst{\ref{inst1}}
\and Wei Zhang (张伟)\inst{\ref{inst1}}\inst{\ref{inst2}}}

\institute{Pengcheng Laboratory, Nanshan District, Shenzhen 518000, China\label{inst1}
\and Harbin Institute of Technology, Shenzhen 518000, China\label{inst2}
}

\date{Received; accepted}

\abstract{
Galaxy image translation refers to a process that maps galaxy images from a source domain to a target domain, which is an important application in galaxy physics and cosmology. With deep learning-based generative models, image translation has been performed for image generation, data quality enhancement, information extraction, and generalized for other tasks such as deblending and anomaly detection. However, most endeavors on image translation primarily focus on the pixel-level and morphology-level statistics of galaxy images. There is a lack of discussion on the preservation of complex high-order galaxy physical information, which would be more challenging but crucial for studies that rely on high-fidelity image translation. Therefore, in our work we investigated the effectiveness of generative models in preserving high-order physical information (represented by spectroscopic redshift) along with pixel-level and morphology-level information. We tested four representative models, i.e. a transformer with shifted windows (Swin Transformer), a super-resolution generative adversarial network (SRGAN), a capsule network, and a diffusion model, performing intra-domain and inter-domain translations using galaxy images from the Sloan Digital Sky Survey (SDSS) and the Canada-France-Hawaii Telescope Legacy Survey (CFHTLS). We found that these models show different levels of incapabilities in retaining redshift information, even if the global structures of galaxies and morphology-level statistics can be roughly reproduced. In particular, the cross-band peak fluxes of galaxies were found to contain meaningful redshift information, whereas they are subject to noticeable uncertainties in the translation of images, which may substantially be due to the nature of many-to-many mapping. Nonetheless, imperfect translated images may still contain a considerable amount of information and thus hold promise for downstream applications for which high image fidelity is not strongly required. Our work can facilitate further research on how complex physical information is manifested on galaxy images, and it provides implications on the development of image translation models for scientific use.
}


\maketitle

\section{Introduction} \label{sec:intro}

Galaxy imaging data are important sources for the studies of galaxy structures, evolution, and cosmology. The next-generation imaging surveys, such as the \textit{Euclid} survey \citep{Laureijs2011}, the \textit{Nancy Grace Roman} Space Telescope \citep{Spergel2015}, the \textit{Vera C. Rubin} Observatory Legacy Survey of Space and Time \citep[LSST;][]{Ivezic2019}, and the China Space Station Telescope \citep[CSST;][]{Zhan2018}, will produce a vast amount of observational galaxy images. Thanks to the rapid development in artificial intelligence (AI), deep learning-based generative models hold great promise in enhancing observational data via image simulation and pushing the boundaries of data-driven research on galaxy physics and cosmology.

Deep learning-based generative models, represented by variational autoencoders \citep[VAEs;][]{Kingma2013}, generative adversarial networks \citep[GANs;][]{Goodfellow2014}, normalizing flows \citep{Rezende2015}, autoregressive models \citep{Germain2015, Uria2016, vandenOord2016, Papamakarios2017, Salimans2017}, transformers \citep{Vaswani2017}, and diffusion models \citep{Ho2020, Song2021, Saharia2022}, can efficiently acquire meaningful information from data and usually generate galaxy images without relying on physical assumptions or modeling. Typically, there are mainly two types of generative models for generating galaxy images. One type of model can ``invent'' new galaxies that do not exist in nature but follow the learned underlying distribution of galaxy properties \citep[e.g.][]{Fussell2019, Dia2020, Spindler2021, Smith2022}. The other type, usually referred to as ``image-to-image translation'' models, can be used to build a mapping between a source image domain and a target image domain. Each image is generated based on the content learned from an input image from the source domain but is expected to inherit the characteristics or prior information from the target domain. With a target domain that has a higher S/N and narrower point spread functions, image translation models can be leveraged to enhance the quality of the images in a source domain by reducing noise, deblurring, and improving image resolution \citep[e.g.][]{Graff2014, Schawinski2017, Buncher2021, Gan2021, Jia2021, Lin2021, Akhaury2022, Li2023, Wang2023, Akhaury2024, Kinakh2024, Miao2024, Park2024, Luo2025, Shan2025}. Such ``inter-domain'' image translation, in which the two domains consist of observed images from different galaxy surveys or simulated paired images (e.g. clean versus corrupted), is a promising way to extend the sky coverage or the size of existing survey data and enhance data quality, which would augment data for image-based applications \citep[e.g.][]{Holzschuh2022} and broaden the extent of scientific discoveries \citep[e.g.][]{Adam2025, Miao2025, Shibuya2025}. On the other hand, when the target domain is identical to the source domain, the ``intra-domain'' translation (usually in the form of autoencoders) can be applied to obtain a low-dimensional representation of the source domain to extract meaningful information for subsequent analysis and understand data structures \citep[e.g.][]{Cheng2021, Zhou2022, Fang2023} or assist with the inter-domain translation \citep[e.g.][]{Lin2021}.

Generalizing the idea of image translation, more complex models and algorithms can be established and implemented for other tasks such as deblending \citep[e.g.][]{Lanusse2019, Reiman2019, Boucaud2020, Arcelin2021, Wang2022, Zhang2024}, anomaly detection \citep[e.g.][]{DAddona2021, StoreyFisher2021}, and imputation of missing data \citep[e.g.][]{Luo2024}. For example, a model that learns the prior for individual galaxies may be extended as a deblender to separate galaxies in blended systems \citep[e.g.][]{Lanusse2019, Arcelin2021}. In a word, deep learning-based image translation is a powerful tool for processing and analyzing large-scale datasets envisioned by future surveys.

Despite its merits, image translation may have difficulties in retaining the physical information of galaxies contained in images. The manifestation of physical information on images may be complex, having high-order dependences on detailed spatial structures and showing high sensitivity to small flux perturbations \citep{Campagne2020, Aleksandra2022}. Hence, it cannot simply be quantified with pixel-based low-order statistics such as integrated flux intensities or galaxy sizes. This is further complicated by the nature of many-to-many mapping intrinsic to image translation. For the inter-domain translation involving two galaxy surveys, the target domain is usually divergent from the source domain with different observational effects and random noise, meaning that the mapping from the source to the target is not unique. Even for the intra-domain translation, there is usually a compression of information (via dimensionality reduction) in a model that results in uncertainties in the translation process. In both cases, the generated images should only be considered as an approximation of the original target images, which would cause a loss of high-order physical information.

We note that retaining complex high-order physical information would potentially be an indispensable part of image translation and crucial for downstream research that requires high-fidelity data enhancement or information extraction. In particular, several studies such as \citet{Zanisi2021} and \citet{Alfonzo2024} have found that spatially resolved detailed morphological structures might be tightly connected with physical properties. For the inter-domain translation, high-fidelity translated images with physical information preserved may reveal fine structures that are unresolved in low-quality data and provide valuable insights into the details of the relations between morphology and physical properties, leading to a deeper understanding of galaxy formation and evolution mechanisms. For the intra-domain translation, the preservation of physical information would produce an informative representation of data suitable for analyzing data structures that involve not only photometric or morphological features but also physical properties. In contrast, a model incapable of retaining physical information would contribute limited scientific significance in these respects.

However, most studies on galaxy image translation and simulation have primarily focused on the reconstruction and evaluation of images at the pixel level or the morphology level, such as flux statistics and galaxy shapes, and they lack discussion on more complex information such as redshift and physical properties. In the mean time, most studies have trained models using pixel-based loss functions such as the mean square error (MSE) and the mean absolute error (MAE), which can force a model to learn low-order flux statistics but may leave high-order features (either single-band or cross-band) unconstrained. Such models are unable to guarantee the preservation of high-order physical information. \citet{LiYunQi2024} applied a conditional denoising diffusion probabilistic model (DDPM) and a conditional VAE for image translation and found that both models resulted in a loss of redshift information, similar to the results obtained by \citet{Lizarraga2024}. While \citet{Ravanbakhsh2016} and \citet{Lanusse2021} input physical information to conditional VAEs in the form of conditions for image simulation, to what extent the physical information can be recovered requires further tests. Additionally, while there are a few metrics that are commonly used for evaluating the quality of natural images, such as the inception score \citep[IS;][]{Salimans2016}, the Fr\'{e}chet inception distance \citep[FID;][]{Heusel2017}, and the structural similarity \citep[SSIM;][]{Wang2004}, these metrics are not fully indicative of the fidelity of galaxy images in terms of high-order physical information. In a word, for optimizing image translation models, the ability to retain high-order physical information remains to be investigated.

\begin{figure}
\centering
\includegraphics[width=1.0\linewidth]{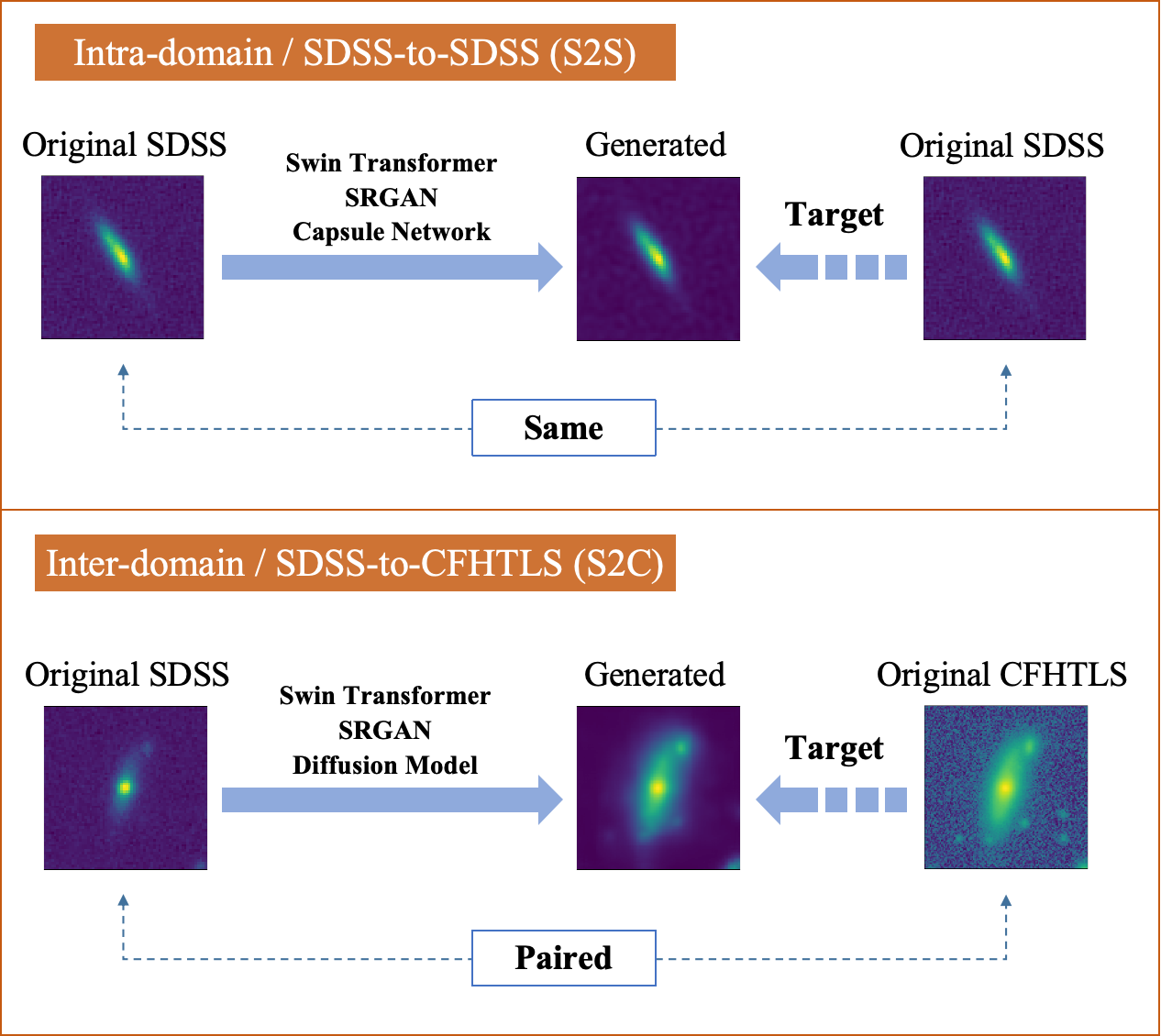}
\caption{Illustration of the intra-domain SDSS-to-SDSS translation (S2S) and the inter-domain SDSS-to-CFHTLS translation (S2C) investigated in this work. The Swin Transformer and the SRGAN were used for both the S2S and S2C translations. The capsule network was used for the S2S translation. The diffusion model was used for the S2C translation.}
\label{fig:s2s_s2c_sketch}
\end{figure}

Therefore, this work is our first attempt toward understanding and preserving high-order physical information in the translation of galaxy images. We tested a few representative generative models for image translation, including a transformer with shifted windows (Swin Transformer), a super-resolution generative adversarial network (SRGAN), a capsule network, and a diffusion model. Using multi-band galaxy images from the Sloan Digital Sky Survey \citep[SDSS;][]{York2000} and the Canada-France-Hawaii Telescope Legacy Survey \citep[CFHTLS;][]{Gwyn2012}, two kinds of translations were performed: an intra-domain translation from SDSS to SDSS itself (denoted as ``S2S'') and an inter-domain translation from SDSS to CFHTLS (denoted as ``S2C''), as shown in Fig.~\ref{fig:s2s_s2c_sketch}. The S2S translation offers a straightforward way to test to what extent the image translation models can retain the original information from the source domain, while the S2C translation was intended to test whether the translated images can match the target domain that is different from the source domain. Via comparing translated images with original target images, we examined the effectiveness of these models in preserving high-order physical information along with pixel-level and morphology-level information. Particularly, spectroscopic redshift has complex connections with multi-band galaxy photometry, morphological features, and physical properties, which offers important implications on the physical processes of galaxies. We thus took spectroscopic redshift as a representative of high-order physical information in our analysis.

Our work offers insights into how complex physical information such as redshift is manifested on galaxy images, points out the inadequacies of current image translation models, and provides implications for their scientific use. Though out of the scope of this work, these results can also provide the basis for the development of image translation models capable of preserving physical information, which would promote scientific discoveries that depend on high-fidelity image translation and establish physically informative data-driven priors for relevant tasks (e.g. deblending).

The organization of this paper is as follows. Section~\ref{sec:data} describes the data used in this work. Section~\ref{sec:methods} introduces our methods, including the image translation models and the techniques for assessing translated images in the aspects of pixel-level information, morphology-level information, and redshift information. Section~\ref{sec:results} presents our results. Section~\ref{sec:conclusion} summarizes this work and gives our conclusion. The details about the network architectures and more results are provided in Appendices~\ref{sec:network} and \ref{sec:res_more}, respectively.

\section{Data} \label{sec:data}

\subsection{SDSS for the S2S translation} \label{sec:data_s2s}

We took the SDSS dataset retrieved by \citet{Pasquet2019} and also used in \citet{Lin2022} and \citet{Lin2024}, which are available from the SDSS Data Release 12 \citep[DR12;][]{Alam2015}. Each galaxy is associated with a spectroscopically measured redshift. Since one of the image translation models examined in our work (i.e. the capsule network) requires morphological classifications, we retrieved morphological class labels from the Galaxy Zoo 1 \citep{Lintott2011}. The Galaxy Zoo galaxies are assigned with de-biased fractions of votes for four classes, i.e. spirals, ellipticals, mergers, and star-and-artefacts. We selected galaxies that have de-reddened $r$-band Petrosian magnitude below 17.8 and spectroscopic redshift below 0.4, and only selected spirals and ellipticals (which we considered as having de-biased fractions of votes for spirals or ellipticals over 0.5). Of these galaxies, we randomly took 40\,000 galaxies as a training sample, and 10\,000 galaxies as a validation sample. These galaxies were ``self-paired'' and used for the S2S translation in our work. The validation sample was used for monitoring the training process of the S2S translation models and also presenting the image translation results in Sect.~\ref{sec:results}. Furthermore, in order to analyze the preservation of redshift information, we took a separate test sample of 103\,305 galaxies with the same selections on $r$-band Petrosian magnitude and spectroscopic redshift but not on morphological classifications, same as those used in \citet{Lin2024}. As implied by the redshift predictions from the Swin Transformer in Sect.~\ref{sec:assredshift}, a training sample of 40\,000 galaxies contains sufficient redshift information and is thus sufficient for developing the S2S translation models in our work.

For each galaxy, there are five stamp images each covering one of the five optical bands $u,g,r,i,z$ and having $64\times64$ pixels in spatial dimensions, with a pixel scale of 0.396 arcsec. The galaxy is placed at the center of each image. The five stamp images of a galaxy form a data instance with $64\times64\times5$ pixels in total, which is to be used for image translation. Since the peak fluxes of different galaxies have a large dynamic range that would hinder the training of the image translation models, we rescaled each image according to the following equation
\begin{equation}\label{eq:norm}
I = \begin{cases}
\ln(I_0+1.0), & I_0>0 \\
-\ln(-I_0+1.0), & I_0<0
\end{cases}
,\end{equation}
where $I$ and $I_0$ stand for the flux intensities of the rescaled and non-rescaled images, respectively, and ``$\ln$'' stands for the natural logarithm with base $e$. Unless otherwise noted, the rescaled images are referred to as ``original'' images in order to differentiate from the images generated by the image translation models.

\subsection{SDSS-CFHTLS pairs for the S2C translation}

In order to perform the inter-domain translation, we took data from the CFHTLS-Wide survey to pair with the SDSS data, regarding the SDSS images as a source domain and the CFHTLS images as a target domain. We did not apply the selections on $r$-band Petrosian magnitude and morphological classifications as in Sect.~\ref{sec:data_s2s}, and used \texttt{ra} and \texttt{dec} of each galaxy from the SDSS dataset to search for a counterpart on the CFHTLS-Wide field images retrieved from \citet{CFHTLST07}. This process resulted in a total of 6573 pairs. The CFHTLS stamp images were made to have $128\times128$ pixels in spatial dimensions (two times the resolution of the SDSS images) to facilitate the development of the S2C translation models. Since the pixel scale of a CFHTLS stamp image is 0.187 arcsec, the full scale over 64 pixels on an SDSS image is slightly wider the scale over 128 pixels on a CFHTLS image. Similar to the SDSS images, the CFHTLS stamp images cover the five optical bands $u,g,r,i,z$ (with $128\times128\times5$ pixels for each data instance) and have a galaxy at the center. They were also rescaled using Eq.~\ref{eq:norm}. From the 6573 SDSS-CFHTLS paired images, we randomly selected 5258 pairs as a training sample (about 80\%), and 1315 pairs as a validation sample (about $20\%$). Likewise, the validation image pairs were used for monitoring the training of the models and presenting the image translation results in Sect.~\ref{sec:results}.

\section{Methods} \label{sec:methods}

In this work, we performed the intra-domain S2S translation and the inter-domain S2C translation with four representative generative models, a transformer with shifted windows \citep[Swin Transformer;][]{SwinTransformer, 2021swinir}, a super-resolution generative adversarial network \citep[SRGAN;][]{Ledig2017}, a capsule network \citep{Sabour2017, Dey2022}, and a diffusion model \citep{Ho2020, Song2021, Saharia2022}. The Swin Transformer and the SRGAN were used for both the S2S and S2C translations. The capsule network was used for the S2S translation. The diffusion model was used for the S2C translation. These models were all trained without providing any prior physical information. The two kinds of translations, S2S and S2C, were performed to investigate how the models behave when the target domain and the source domain are identical or have unmatched characteristics. In order to analyze the fidelity of images generated by the translation models, we conducted assessments on pixel-level information, morphology-level information, and redshift information. In particular, spectroscopic redshift is considered to represent high-order physical information. The models and the assessment techniques are detailed in the following subsections, respectively. The network architectures of all the models are presented in Appendix~\ref{sec:network}.

\subsection{Models}\label{subsec:models}

\subsubsection{Swin Transformer}

The Swin Transformer \citep{SwinTransformer} was chosen in our work, as it has a transformer architecture mainly featured by shifted windows and multi-head attention layers, and quite different from the more traditional convolutional neural networks (CNNs). We wanted to investigate whether these characteristics would be helpful for image translation. We adopted the Swin Transformer model from \citet{2021swinir} that was proposed for image super-resolution, which consists of a shallow feature extraction module, residual Swin Transformer blocks (RSTBs), and a high-quality (HQ) image reconstruction module. Based on this model\footnote{Code obtained from https://github.com/cszn/KAIR}, we made two versions to be used for the S2S and S2C translations, respectively. There is a major difference between the two versions. For the S2C translation where the SDSS images are the input and the CFHTLS counterparts are the target, a pixel shuffle layer was applied in the HQ image reconstruction module to upscale the images from $64 \times64$ to $128\times128$ resolution, and there was a skip connection that directly links the shallow feature extraction module to the HQ image reconstruction module. However, as the target domain is identical to the source domain in the S2S translation, in order to prevent the model from copying the input to produce the output, we discarded the skip connection above and applied a $4\times4$ average pooling layer for downsampling in the HQ image reconstruction module, accompanied with two pixel shuffle layers for upsampling. The total number of trainable parameters is about 0.6 million for the S2S translation and 0.5 million for the S2C translation. 

In the training process, an MSE loss was used to contrast the output images and the target images. The mini-batch size was set to 16. We kept the training until the loss converged at 100 training epochs for the S2S translation, and 45 epochs for the S2C translation.

\subsubsection{SRGAN}

The SRGAN \citep{Ledig2017} is a GAN applied for image super-resolution. A typical GAN consists of a generator and a discriminator, which are trained alternately in a generative adversarial training process. The generator is trained to produce fake images indistinguishable from the real ones, and the discriminator is trained to tell apart the fake and real images. We intended to check whether such generative adversarial training process would be beneficial for extracting information from data beyond low-order statistics, particularly redshift information. Based on the SRGAN model from \citet{2022basicsr}\footnote{Code obtained from https://github.com/XPixelGroup/BasicSR}, we made two versions for the S2S and S2C translations, respectively, similar to the Swin Transformer. We applied two $2\times2$ average pooling layers accompanied with two pixel shuffle layers in the generator for the S2S translation, while a pixel shuffle layer was applied for the S2C translation. The total number of trainable parameters is about 9.1 million for the S2S translation and 15.5 million for the S2C translation.

In the training process, an MSE loss was used to contrast the output images from the generator and the target images, and a cross-entropy adversarial loss with binary classification was applied for discriminating the generated and original target images. Same as the Swin Transformer, we conducted 100 and 45 training epochs for the S2S and S2C translations, respectively.

\subsubsection{Capsule network}

The capsule network \citep{Sabour2017} is a special kind of deep learning model, whose building blocks are capsules (composed of a group of neurons). It is robust to rotations and viewpoints and may have a better ability to learn a low-dimensional representation of galaxies than conventional CNNs. We modified the model from \citet{Dey2022}\footnote{Code obtained from https://biprateep.github.io/encapZulate-1}. In particular, we implemented an encoder built with capsule modules to extract information from the input images and output two low-dimensional vectors, each having 16 nodes. The two vectors correspond to the two morphological classes (i.e. spirals and ellipticals). During training, the Euclidean length of the vector corresponding to the morphological class label was maximized with a margin loss. This vector was input to a decoder to produce output images, contrasted with the target images using an MSE loss. The Euclidean length of the other vector was minimized simultaneously with the margin loss. During inference, the vector with larger Euclidean length was chosen as the input to the decoder. In other words, the image translation by the capsule network leverages extra information from the morphological classifications. As the capsule network was only used for the S2S translation, the target and the input images are the same SDSS images.
 
The total number of trainable parameters of the capsule network is about 6.4 million. In the training process, the mini-batch size was set to 16. The loss converged at 100 training epochs. As we intended to investigate whether the morphological classifications would be helpful for the preservation of physical information, we also implemented a modified version that was the same as our default capsule network implementation except that only one vector was output from the encoder and the morphological classifications were not used.

\subsubsection{Diffusion model}

We chose the diffusion model as it is a probabilistic model and may have the potential to reproduce detailed features without generative adversarial training. For the S2C translation, we modified the model leveraged by \citet{Kinakh2024}\footnote{Code obtained from https://github.com/vkinakh/Hubble-meets-Webb}, a conditional version \citep{Saharia2022} integrated with the principles of the denoising diffusion implicit model \citep[DDIM;][]{Song2021}. There are typically two processes in this diffusion model. The forward process corrupts the target CFHTLS images with varying levels of Gaussian noise, while the generative process, defined as the reverse of the forward process, recovers denoised images in an iterative manner using a neural network conditioned on the original SDSS counterparts. Different from the denoising diffusion probabilistic models \citep[DDPMs;][]{Ho2020} of which the generative process and the forward process are forced to have the same number of iterative steps (denoted as $T$), the DDIM accelerates the generative process by regarding it as a non-Markovian process and taking a sub-sequence with much smaller length from the full sequence of the $T$ diffusion steps. In this work, we took a sub-sequence of 100 steps with equal spacing from a total of $T=1000$ steps (i.e. one step reserved from every ten steps). A U-Net architecture \citep{Ronneberger2015} was applied to realize the generative process, where the corrupted images at a given diffusion step and the original SDSS images (as conditions) were both fed into the network to predict noise images that were then used to construct the corrupted images at the previous diffusion step in the sub-sequence, eventually leading to output images that approximate the CFHTLS target images. The total number of trainable parameters of the U-Net is about 62.6 million. In the training process, the mini-batch size was set to 16. An MSE loss was leveraged to contrast the predicted noise images and the actual noise images that were used to corrupt the target CFHTLS images at a given step, and the loss converged at 2000 training epochs.

Since there is significant stochasticity in the diffusion model, the model controllability has to be improved. Rather than applying pure Gaussian noise images as the starting point in the reverse process, we first applied the forward process to produce noise images using the fake CFHTLS images translated from the original SDSS images by the Swin Transformer, then used such noise images in the reverse process to generate denoised images. Because the noise images contain information from the Swin Transformer-generated images, this operation was found to make the image generation process of the diffusion model much less unstable. We also note that there was no leakage of any information about the target domain to the model, because the original CFHTLS images were never used as the input.

\subsection{Assessment techniques}\label{subsec:ass_tech}

\subsubsection{Pixel-level information}\label{subsubsec:pixinfo}

To assess the pixel-level information, we considered the mean fluxes (rescaled with Eq.~\ref{eq:norm}) estimated over the full image and over the four central pixels in a certain band for each galaxy. They are defined as
\begin{equation}\label{eq:diffwhole}
f_{full} = \frac{1}{N_f} \sum_{i=0}^{N_f} f_{i},
\end{equation}
and
\begin{equation}\label{eq:difffour}
f_{central} = \frac{1}{N_c} \sum_{j=0}^{N_c} f_{j},
\end{equation}
where $i$ runs over all pixels of an image; $j$ runs over the four central pixels; $N_f$ stands for the total number of pixels; $N_c$ equals 4. $f_{full}$ characterizes the mean flux level contributed by the central galaxy, the background and potentially other sources, while $f_{central}$ describes the peak flux of the galaxy. These two fluxes have different physical meanings.

\subsubsection{Morphology-level information}\label{subsubsec:objinfo}

To characterize the morphology-level information, we studied five representative galaxy properties estimated using \texttt{Photutils}\footnote{https://photutils.readthedocs.io/en/stable/index.html} \citep{larry_bradley_2024_13989456}. These properties are as follows.
\begin{itemize}
\item The scale of the semimajor axis, $a$, defined as the $1\sigma$ dispersion along the semimajor axis.

\item The ellipticity, $e$, defined as
\begin{equation}\label{eq:ellipticity}
e = 1-\frac{a}{b},
\end{equation}
where $a$ and $b$ are the semimajor and semiminor axes, respectively.

\item The AB magnitude based on the Kron flux, i.e.
\begin{equation}\label{eq:magnitude}
Mag = \begin{cases}
22.5 - 2.5\log f_{\mathrm{Kron}} & (\mathrm{SDSS}) \\
30 - 2.5 \log f_{\mathrm{Kron}} & (\mathrm{CFHTLS})
\end{cases}
.\end{equation}

\item The circularized full width at half maximum (FWHM) of the two-dimensional profile of a galaxy.

\item The half-light radius, $R_{50}$, defined as the circular radius that encloses 50\% of the Kron flux.
\end{itemize}

Before estimating these properties, all the images were restored to the initial flux scale using the reverse of Eq.~\ref{eq:norm}.

\subsubsection{Redshift information}

The fluxes or galaxy properties listed in Sects.~\ref{subsubsec:pixinfo} and \ref{subsubsec:objinfo} cannot reveal whether the image translation models have the ability to preserve complex and high-order physical information. In particular, spectroscopic redshift is a good representative of such high-order physical information, and one of the most important quantities in the studies of galaxy physics and cosmology. Therefore, we tested how well redshift information is preserved in the images generated by the translation models. 

We adopted the ``vanilla CNN'' version of the redshift estimation models from \citet{Lin2024}, in which an encoder projects multi-band galaxy images and galactic reddening $E(B-V)$ obtained from \citet{Schlegel1998} to a low-dimensional representation, and an estimator takes the representation as its input to produce a probability density estimate of redshift and the probability-weighted mean is taken as the redshift point estimate. To test the deviations between the generated and original images with respect to the features that encode redshift information, we used redshift estimation models trained with original SDSS or CFHTLS images to predict redshifts for the images generated by the S2S or S2C translations, respectively (denoted as ``original-generated''). The generated images were translated via either S2S or S2C from the original images in the SDSS test sample, different from those used for training the redshift estimation models (denoted as the ``$z$-training'' data to avoid confusion with the training samples for image translation). The generated CFHTLS images were cropped to $64\times64\times5$ to feed into the redshift estimation model for CFHTLS. Furthermore, to investigate to what extent the image translation process can preserve redshift information regardless of how it is manifested on the generated images, we predicted redshifts for the same generated images from the test sample but using models trained with generated fake SDSS or CFHTLS images, which were translated from the original SDSS $z$-training data via each translation model, separately (denoted as ``generated-generated''). These two cases were both compared with the baseline in which a redshift estimation model trained with the original SDSS $z$-training images was used to make predictions for the original SDSS test images. The details of the training procedures and the SDSS and CFHTLS $z$-training data can be found in \citet{Lin2024}.

\begin{figure*}
\begin{center}
\centerline{\includegraphics[width=0.75\linewidth]{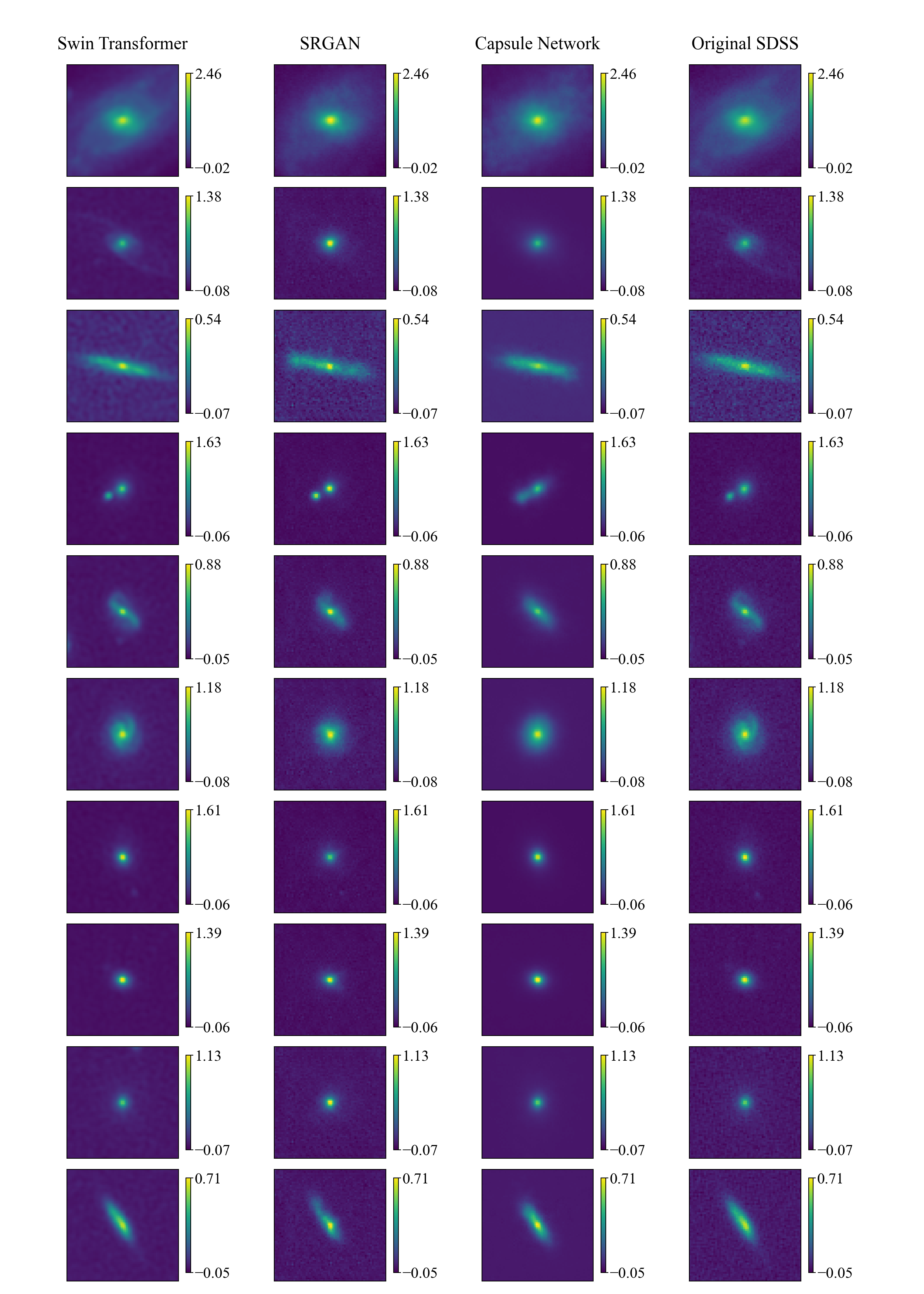}}
\caption{Exemplar $r$-band images generated by the S2S translation models (i.e. the Swin Transformer, the SRGAN, and the capsule network) and the corresponding original SDSS validation images. The color bar next to each image indicates the rescaled flux (Eq.~\ref{eq:norm}) in the natural logarithmic scale.}
\label{fig:s2s_10samps}
\end{center}
\end{figure*}

\begin{figure*}
\begin{center}
\centerline{\includegraphics[width=0.85\linewidth]{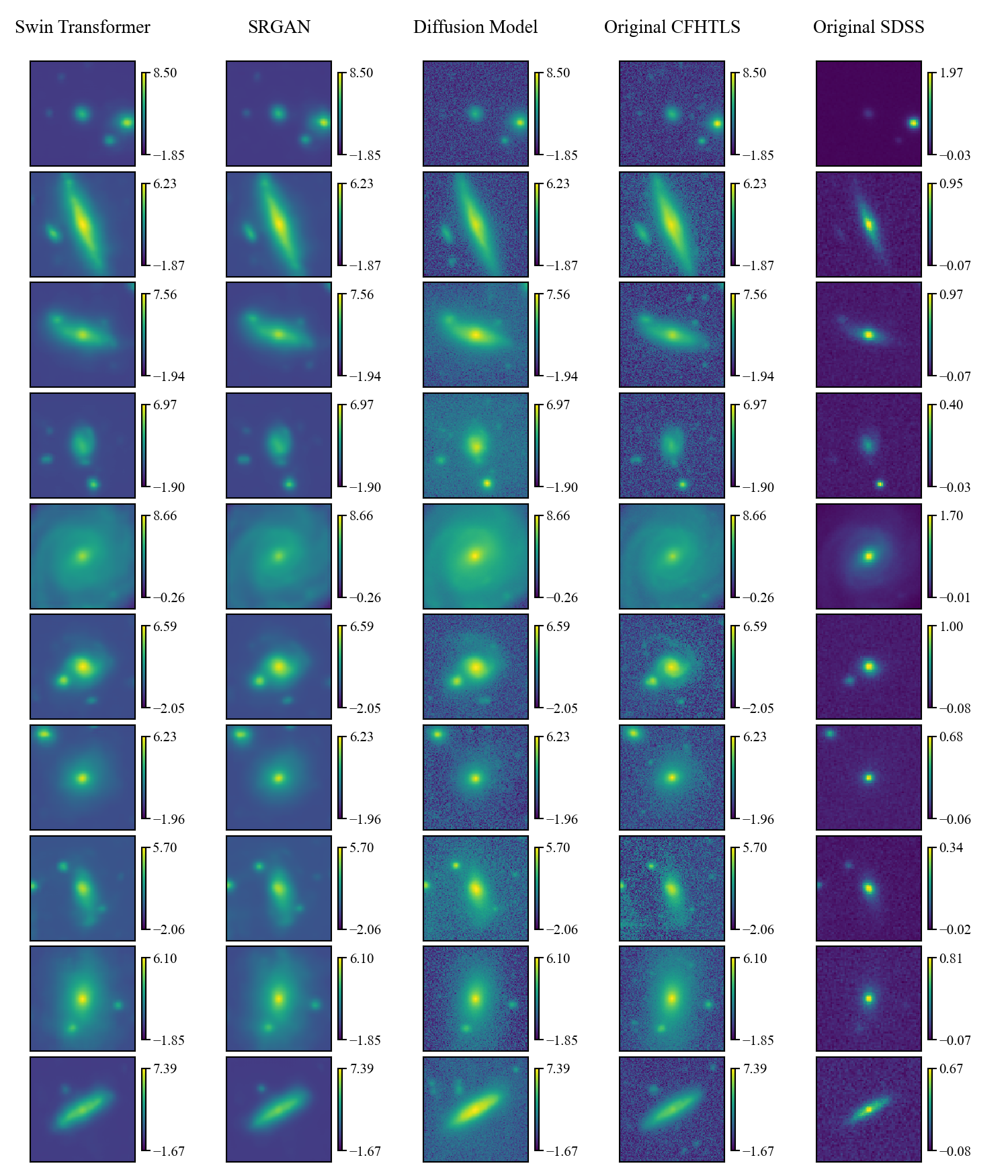}}
\caption{Exemplar $r$-band images generated by the S2C translation models (i.e. the Swin Transformer, the SRGAN, and the diffusion model) and the corresponding original CFHTLS validation images (target) and original SDSS validation images (input). The color bar next to each image indicates the rescaled flux (Eq.~\ref{eq:norm}) in the natural logarithmic scale.}
\label{fig:s2c_10samps}
\end{center}
\end{figure*}

\begin{figure*}
\begin{center}
\centerline{\includegraphics[width=1.0\linewidth]{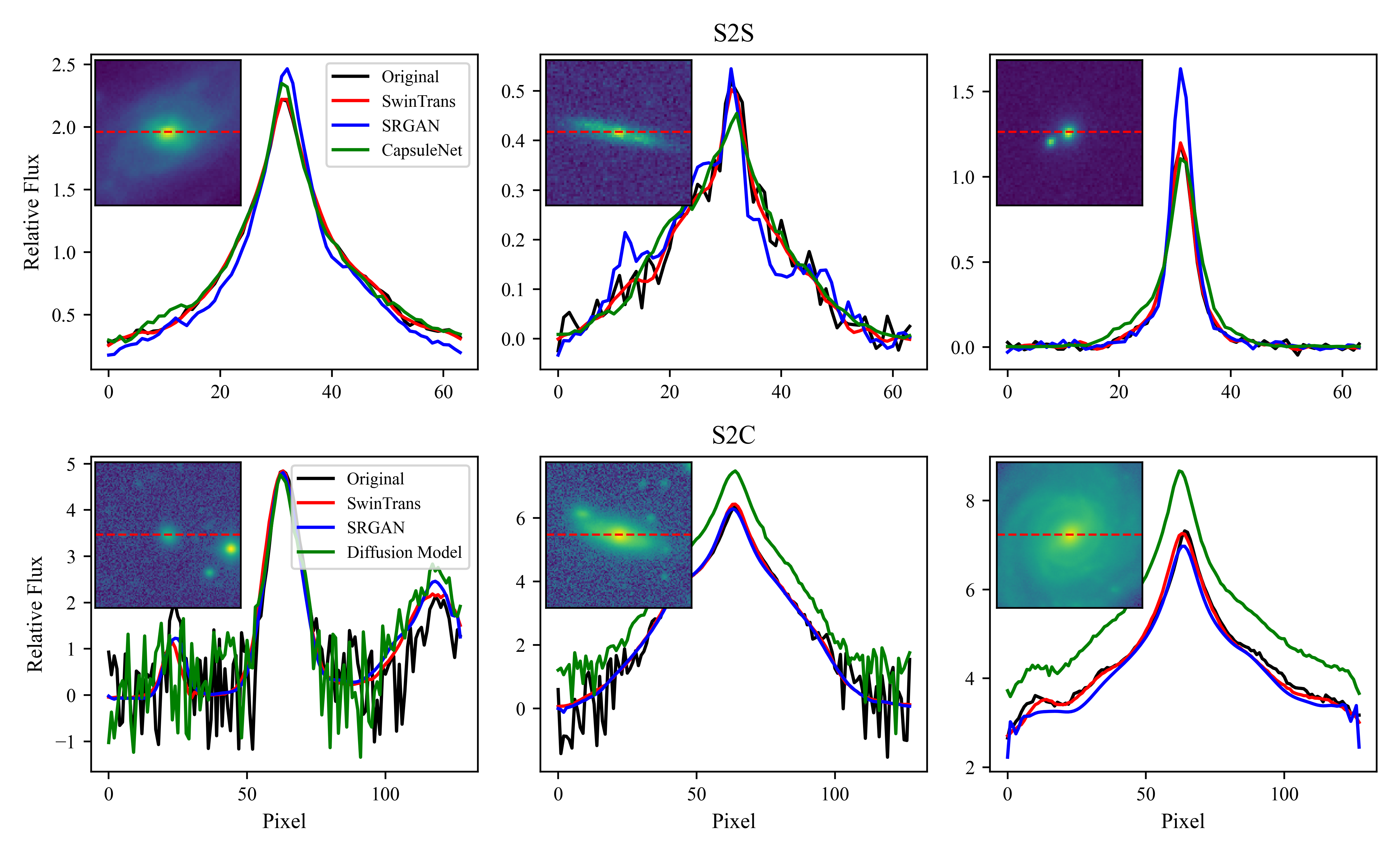}}
\caption{Flux profiles of exemplar S2S and S2C translated $r$-band images compared to the original SDSS and CFHTLS validation images (also illustrated in Figs.~\ref{fig:s2s_10samps} and \ref{fig:s2c_10samps}), respectively. The fluxes (rescaled with Eq.~\ref{eq:norm}) are shown in the natural logarithmic scale. \textit{First row:} Flux profiles from the S2S translation corresponding to the original SDSS validation images and the counterparts generated by the Swin Transformer, the SRGAN, and the capsule network, respectively. The upper-left corner of each panel shows the exemplar original $r$-band image, where the red dashed line through the image center marks the pixels from which the flux profiles are drawn. \textit{Second row:} Flux profiles from the S2C translation corresponding to the original CFHTLS validation images and the counterparts generated by the Swin Transformer, the SRGAN, and the diffusion model, respectively.}
\label{fig:s2s_s2c_flux_profile}
\end{center}
\end{figure*}

\section{Results} \label{sec:results}

\subsection{Assessment on pixel-level information} \label{sec:asspixel}

\begin{figure*}
\begin{center}
\centerline{\includegraphics[width=1.0\linewidth]{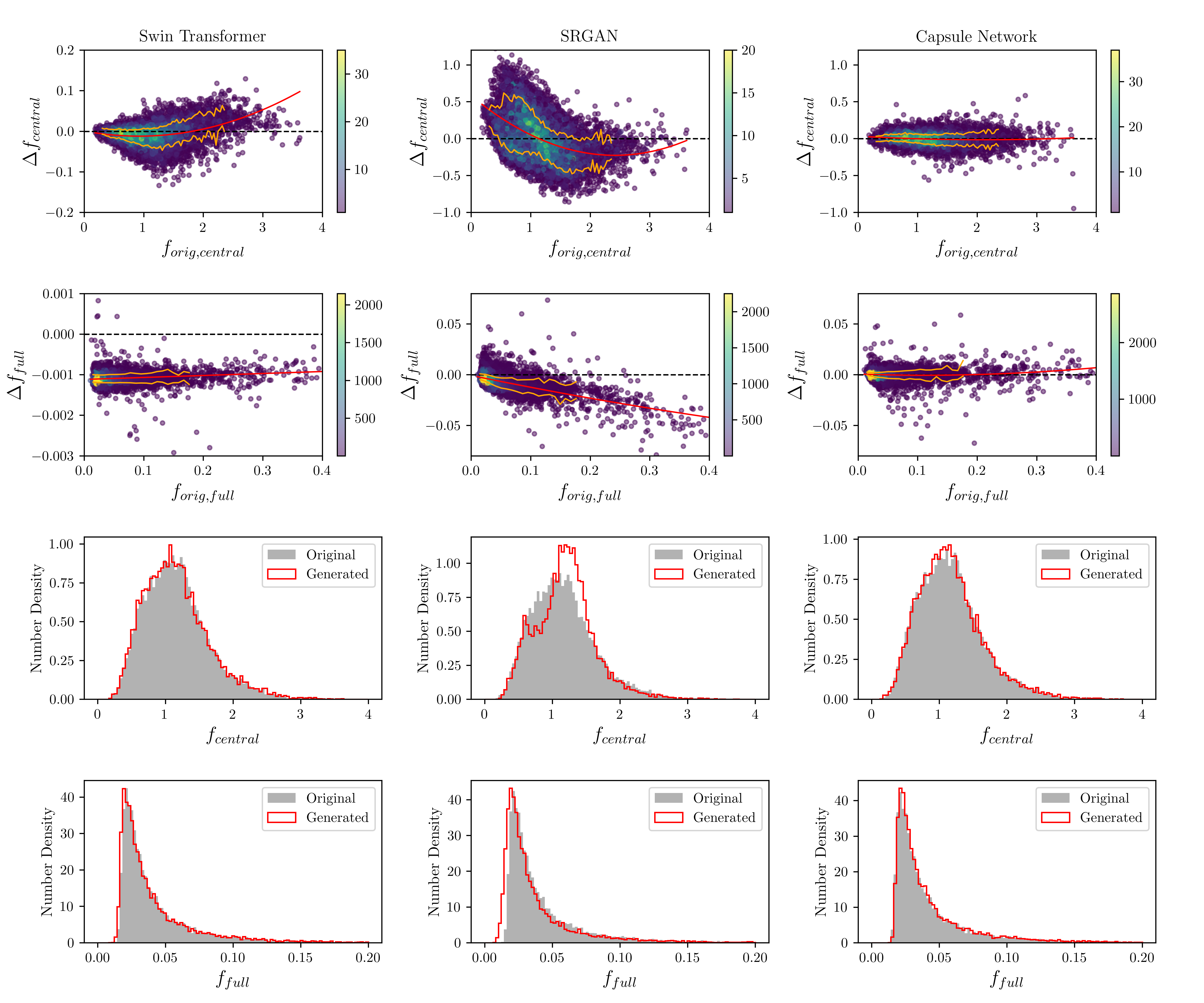}}
\caption{Comparison on flux intensities between the S2S translated $r$-band images and original SDSS validation images. The fluxes (rescaled with Eq.~\ref{eq:norm}) are shown in the natural logarithmic scale. \textit{First row:} Mean central flux residuals (estimated over the four central pixels) between the generated and original $r$-band images as a function of the mean central fluxes of the original images, shown for the Swin Transformer, the SRGAN, and the capsule network. In each panel, the orange curves indicate the 16th and 84th percentiles of the flux residuals at each flux level, the red curve indicates the linear or quadratic fit, and the black dashed line indicates zero residuals. All the data points are color coded with the relative number density. \textit{Second row:} Same as the first row, but with the mean fluxes over the full $r$-band images. \textit{Third row:} Mean central flux distributions for the generated and original images. \textit{Fourth row:} Same as the third row, but with the mean full-image fluxes.}
\label{fig:s2s_flux}
\end{center}
\end{figure*}

\begin{figure*}
\begin{center}
\centerline{\includegraphics[width=1.0\linewidth]{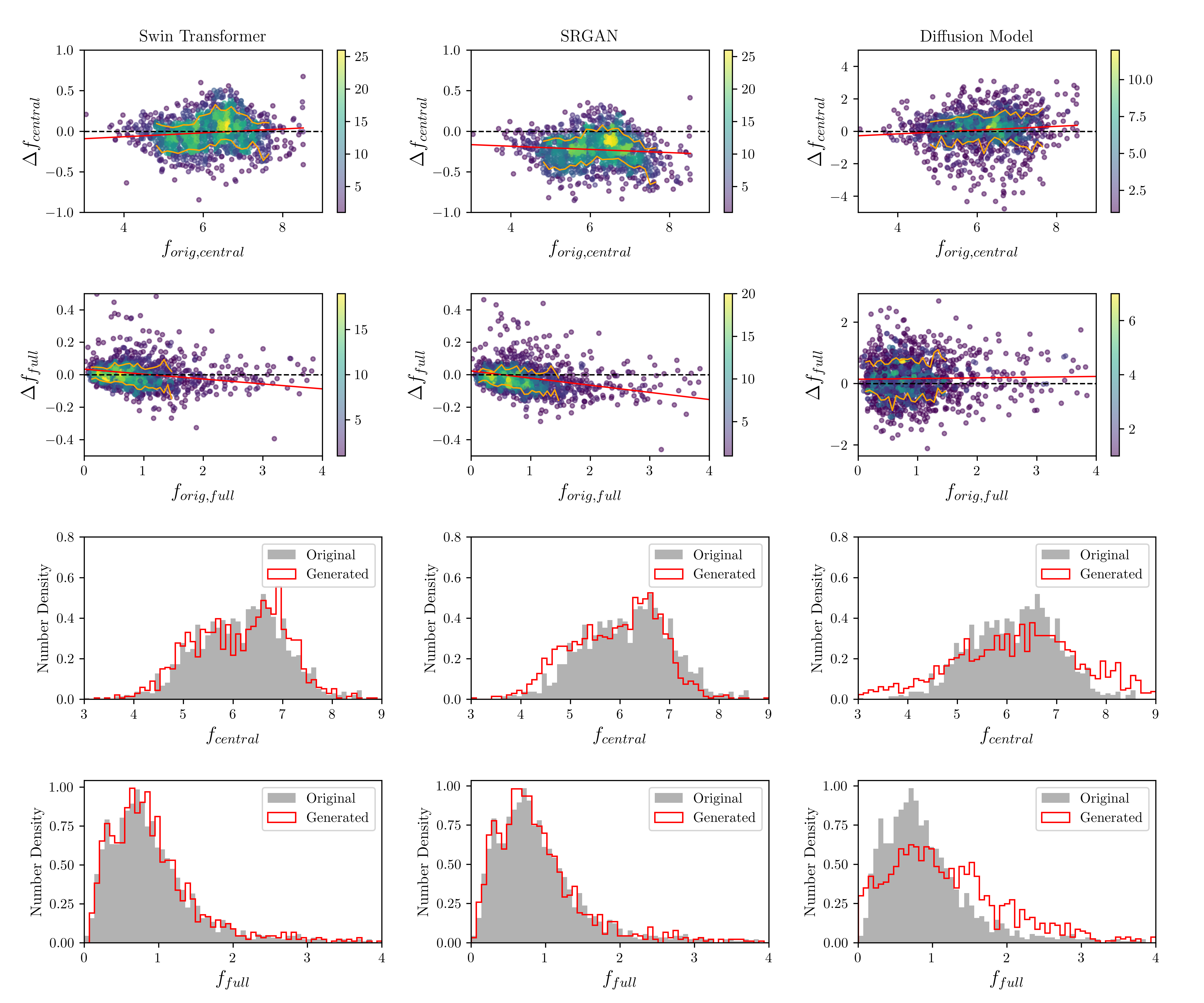}}
\caption{Same as Fig.~\ref{fig:s2s_flux}, but shown for the S2C translated $r$-band images and original CFHTLS validation images.}
\label{fig:s2c_flux}
\end{center}
\end{figure*}

For the assessment on the pixel-level information, we show a few exemplar $r$-band images generated by the S2S and S2C translations in Figs.~\ref{fig:s2s_10samps} and \ref{fig:s2c_10samps}, respectively, compared to the original SDSS or CFHTLS validation images. The RGB versions of these images are presented in Appendix~\ref{sec:res_more}. We also compare the flux profiles of a few generated and original $r$-band images in Fig.~\ref{fig:s2s_s2c_flux_profile}, where each flux profile is drawn along a row that goes through the image center. At first glance, all the models can generate images and flux profiles that are generally comparable with the original ones. They can approximately reproduce both the central galaxies and other sources on the images, and enhance the weak signals not easily recognizable by eyes on the SDSS images when translating to the CFHTLS counterparts.

Among the models, the Swin Transformer performs the best in reconstructing the global structures of galaxies, especially reproducing the peak fluxes. The capsule network is less competent in this regard probably because it undergoes a too heavy reduction of degrees of freedom to bear enough information (from the original $64\times64\times5$ dimensions to a vector of length 16). In fact, the capsule network faces more challenges in reconstructing detailed features than all the other models. While the Swin Transformer may copy some level of flux fluctuations (i.e. noise) directly from the input images in the S2S translation, the capsule network only produces smoothed images as detailed information has essentially been lost. The SRGAN may generate some fluctuation patterns (though not stochastic) in the S2S translation due to the use of generative adversarial training, whereas for the very same reason it would have difficulties in constraining the peak fluxes. Both the Swin Transformer and the SRGAN fail to generate flux fluctuations similar to the original CFHTLS images in the S2C translation due to a lack of stochasticity for producing random noise. On the other hand, the diffusion model can reproduce flux fluctuations that seemingly mimic the noise characteristics thanks to stochasticity, but it suffers from a high level of instability that may result in significant deviations between the generated and original images, even after using the Swin Transformer-generated images to produce noise images in the forward process.

We then analyzed the mean fluxes estimated over the four central pixels and over the full image using the S2S and S2C translated $r$-band images compared to the original validation images, illustrated in Figs.~\ref{fig:s2s_flux} and \ref{fig:s2c_flux}. Though not shown, similar behaviors can be found in other bands. It is noticeable that the Swin Transformer reaches a good control of both the mean central flux residuals $\Delta f_{central}$ (i.e. differences between the generated and original fluxes) and the mean full-image flux residuals $\Delta f_{full}$. In both the S2S and S2C translations, $\Delta f_{central}$ and $\Delta f_{full}$ produced by the Swin Transformer are closer to zero in comparison with the other models, and do not vary drastically as a function of the mean original central fluxes $f_{orig,central}$ or the mean original full-image fluxes $f_{orig,full}$.

Nonetheless, for all the models, $\Delta f_{central}$ typically exhibit a larger range of variation than $\Delta f_{full}$, indicating that the determination of the central flux of a galaxy tends to be more uncertain than the determination of its global shape and the background. In fact, such uncertainties would be unfavorable for the preservation of redshift information and undermine the fidelity of a generated galaxy image, because there appear to be intricate trends between the central fluxes and redshift, as discussed in Sect.~\ref{sec:assredshift}. 

Furthermore, there exist non-zero dependences of $\Delta f_{central}$ ($\Delta f_{full}$) on $f_{orig,central}$ ($f_{orig,full}$) especially for the SRGAN in the S2S translation, which may be due to mixed effects. In particular, attenuation biases would be introduced when there is a lack of information conveyed to the generated images compared to that contained in the original target images, which may be due to information compression in the S2S translation or divergent characteristics associated with the two domains in the S2C translation. That is, the regression coefficients of the quantities derived from the generated images on those derived from the original counterparts would be attenuated toward zero, resulting in a shrunk dynamic range. Hence, such attenuation biases would make $\Delta f_{central}$ ($\Delta f_{full}$) globally overestimated at low $f_{orig,central}$ ($f_{orig,full}$) values while underestimated at high $f_{orig,central}$ ($f_{orig,full}$) values \citep{Ting2025}. Other biases in determining the flux profiles would also project to a shift in $\Delta f_{central}$ ($\Delta f_{full}$) and may reverse the decreasing slopes of $\Delta f_{central}$ ($\Delta f_{full}$) as a function of $f_{orig,central}$ ($f_{orig,full}$). It is also noteworthy that the variation of $\Delta f_{full}$ in the S2S translation is at least an order of magnitude lower than that of $\Delta f_{central}$ but much higher in the S2C translation, which may be due to a shortage of SDSS-CFHTLS paired training data and that the SDSS and CFHTLS images have unmatched noise. The errors in calibrating the mean fluxes and the nonflat slopes reflect the dispersions and biases inherited from the nature of many-to-many mapping for image translation.

For the Swin Transformer and the capsule network, there is good consistency between the mean flux distributions for the generated and original images, though the capsule network suffers from larger errors in determining the mean fluxes. In contrast, the SRGAN-generated $f_{central}$ distribution in the S2S translation noticeably deviates from the original distribution. The left and right peaks in this $f_{central}$ distribution roughly correspond to star-forming and passive galaxies in the SDSS sample, respectively. The presence of the two peaks may originate from the use of generative adversarial training, which makes the SRGAN overly learn the features of over-represented galaxies and aggravate attenuation biases within each galaxy population. Such biases also lead to a skewed slope of $\Delta f_{central}$ as a function of $f_{orig,central}$. Lastly, the diffusion model produces much wider mean flux distributions due to larger flux errors.

To summarize, the Swin Transformer outperforms the other models in reconstructing the global structures of galaxies and constraining the flux errors and biases, especially reproducing the peak fluxes that contain physically meaningful information. We note that this does not necessarily mean that a transformer architecture is universally superior over other architectures for image translation, but that a model should have a sufficient number of effective degrees of freedom with high flexibility in order to properly establish a mapping between two image domains. The Swin Transformer, the SRGAN and the capsule network all lack a stochastic mechanism to produce the random noise background that resembles the original images. The lack of stochasticity in an image translation model may be considered for denoising images. Although the smoothed images, generated by the Swin Transformer and the SRGAN in the S2C translation and the capsule network in the S2S translation, are seemingly preferable due to a reduction of noise, they should be taken with a grain of salt because such smoothing would potentially alter real signals or erase detailed features, jeopardizing the preservation of physical information. Furthermore, while the diffusion model may reproduce flux fluctuations that mimic the noise patterns, it is less controllable and results in larger flux errors. The SRGAN also has some level of instability and may suffer from enlarged attenuation biases and flux errors. This underscores the importance of model stability for generating images with good fidelity.

\subsection{Assessment on morphology-level information} \label{sec:assphypro}

\begin{figure*}
\begin{center}
\centerline{\includegraphics[width=1.0\linewidth]{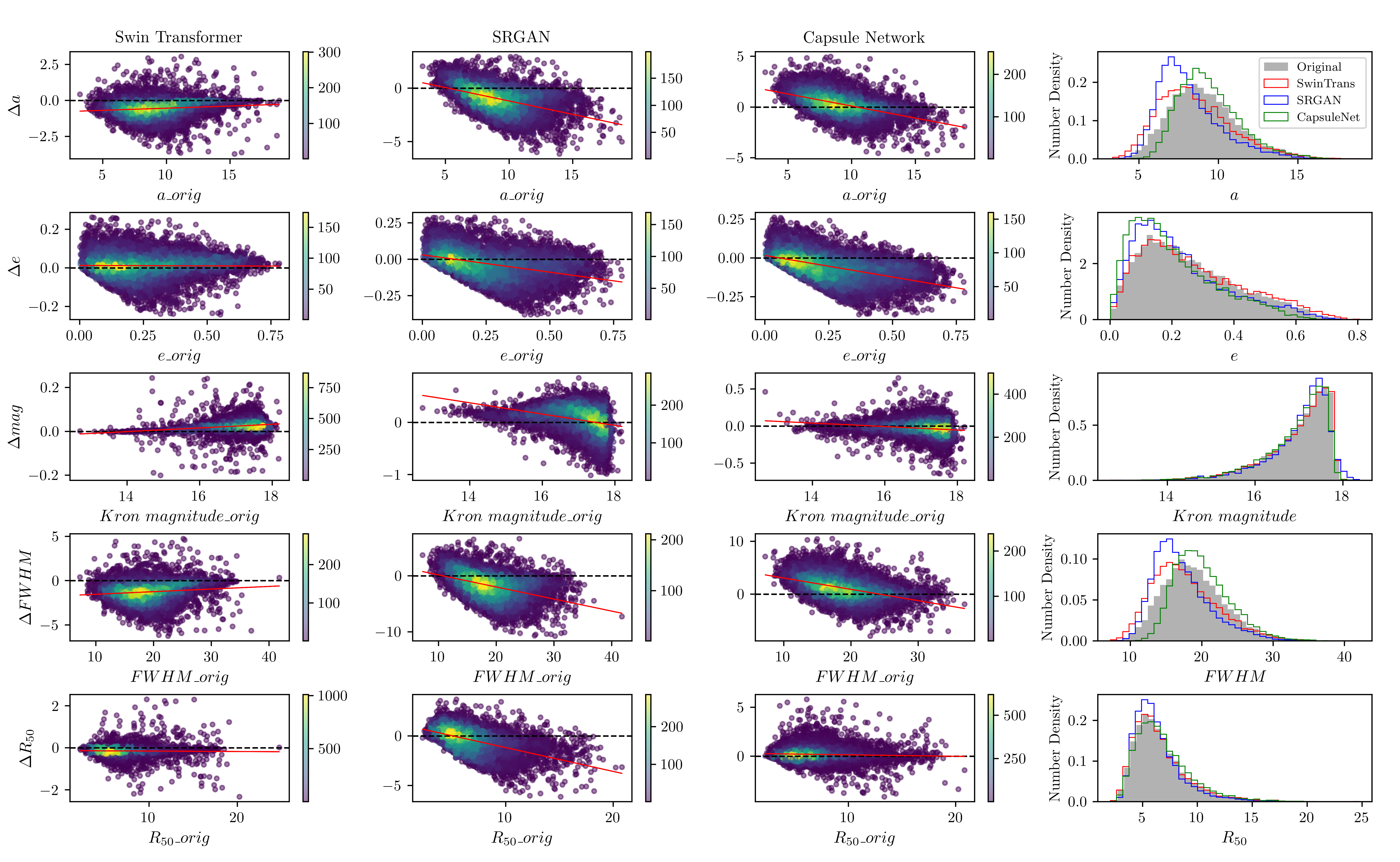}}
\caption{Comparison on galaxy properties estimated using the S2S translated $r$-band images and original SDSS validation images. The shown galaxy properties include the semimajor axis, $a$; the ellipticity, $e$; the AB magnitude based on the Kron flux; the FWHM; and the half-light radius, $R_{50}$. \textit{First three columns:} Residuals of each galaxy property between the generated and original $r$-band images as a function of the values estimated using the original images, shown for the Swin Transformer, the SRGAN, and the capsule network. In each panel, the red line indicates the linear fit, and the black dashed line indicates zero residuals. The outliers on the $y$ axis are removed in order to improve the visibility of the main trends. All the shown data points are color coded with the relative number density. \textit{Fourth column:} Distributions of each galaxy property for the generated and original images.}
\label{fig:s2s_prop}
\end{center}
\end{figure*}

\begin{figure*}
\begin{center}
\centerline{\includegraphics[width=1.0\linewidth]{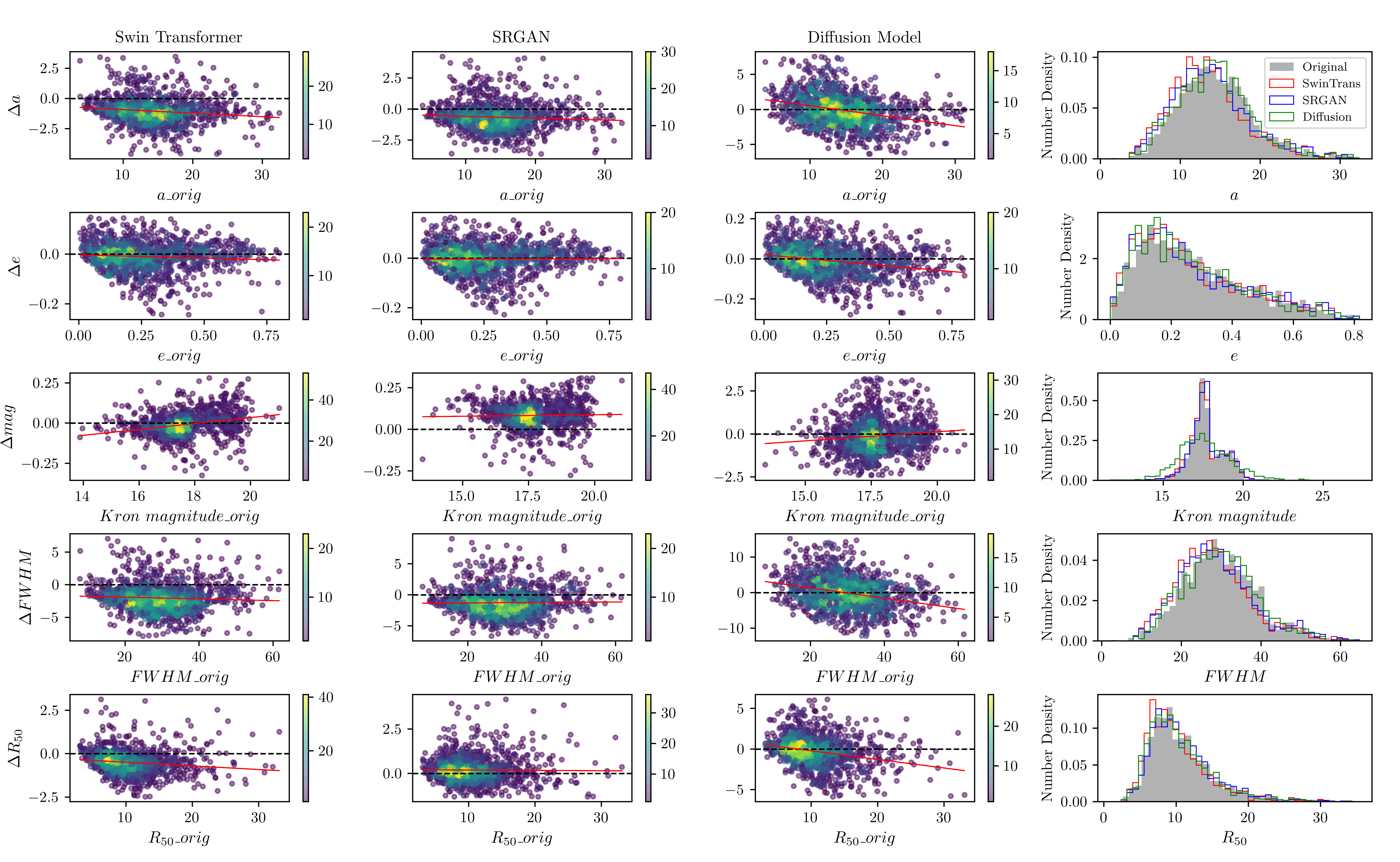}}
\caption{Same as Fig.~\ref{fig:s2s_prop}, but shown for the S2C translated $r$-band images and original CFHTLS validation images.}
\label{fig:s2c_prop}
\end{center}
\end{figure*}

For the assessment on morphology-level information, Figs.~\ref{fig:s2s_prop} and \ref{fig:s2c_prop} illustrate the residuals and the distributions of five galaxy properties estimated using the S2S and S2C translated $r$-band images with reference to the original validation images. These properties are the semimajor axis, $a$; the ellipticity, $e$; the AB magnitude based on the Kron flux; the FWHM; and the half-light radius, $R_{50}$ (defined in Sect.~\ref{subsubsec:objinfo}). The results obtained with other bands are similar to those with $r$ band. Overall, the Swin Transformer has the best performance in constraining the residuals and approximating the original distributions of all the properties, demonstrating its ability in preserving morphology-level information. Despite this, all the models cannot circumvent dispersions for any of the five properties shown in the residual plots. The dispersions may be due to the fact that the measurements of the properties using the generated and original images are both subject to uncertainties, which do not necessarily imply that the models are incapable of extracting meaningful morphology-level information. In most residual plots, we also noticed the existence of attenuation biases as discussed in Sect.~\ref{sec:asspixel}. In particular, all the models tend to over-predict small $a$, $e$, FWHM, and $R_{50}$ values, as indicated by the falling lower edge that bounds the data points in each residual plot. In addition, the dispersions for faint galaxies tend to be larger than those for bright galaxies especially in the S2S translation, implying that more care should be taken for constraining the brightness of faint galaxies. Again, the dispersions and biases point to the nature of many-to-many mapping for translating images.

By checking the galaxy property distributions, we found that all the models can roughly retain the morphology-level statistics of the original data. The Swin Transformer accurately reproduces the $e$, Kron magnitude, and $R_{50}$ distributions. The reproduced $a$ and FWHM distributions are slightly shifted leftward relative to the original ones. This is because the different noise levels on the Swin Transformer-generated and original images affect the source detection and thus result in slightly different moment-based estimates such as the semimajor axis $a$ and the FWHM. In this sense, the half-light radius $R_{50}$ is a more robust quantity to characterize the galaxy size. In contrast with the Swin Transformer, the capsule network finds it difficult to produce small $a$, small FWHM, and large $e$ values, and makes their distributions more concentrated than the original ones. This implies that the capsule network tends to produce smeared images and make inclined galaxies rounder. The SRGAN also makes the $a$, $e$, FWHM, and $R_{50}$ distributions more concentrated in the S2S translation, which may be due to the aggravation of attenuation biases as discussed in Sect.~\ref{sec:asspixel}. The diffusion model has difficulties in reproducing the magnitude distribution, even though the distributions of the other properties closely follow the original ones, implying that the diffusion model may properly reproduce the spatial features while the determination of the flux level is unstable. In short, these results highlight the importance of a stable and flexible model such as the Swin Transformer for learning and accurately preserving morphology-level information in image translation.

\subsection{Assessment on redshift information} \label{sec:assredshift}

\begin{figure}
\centering
\includegraphics[width=1.0\linewidth]{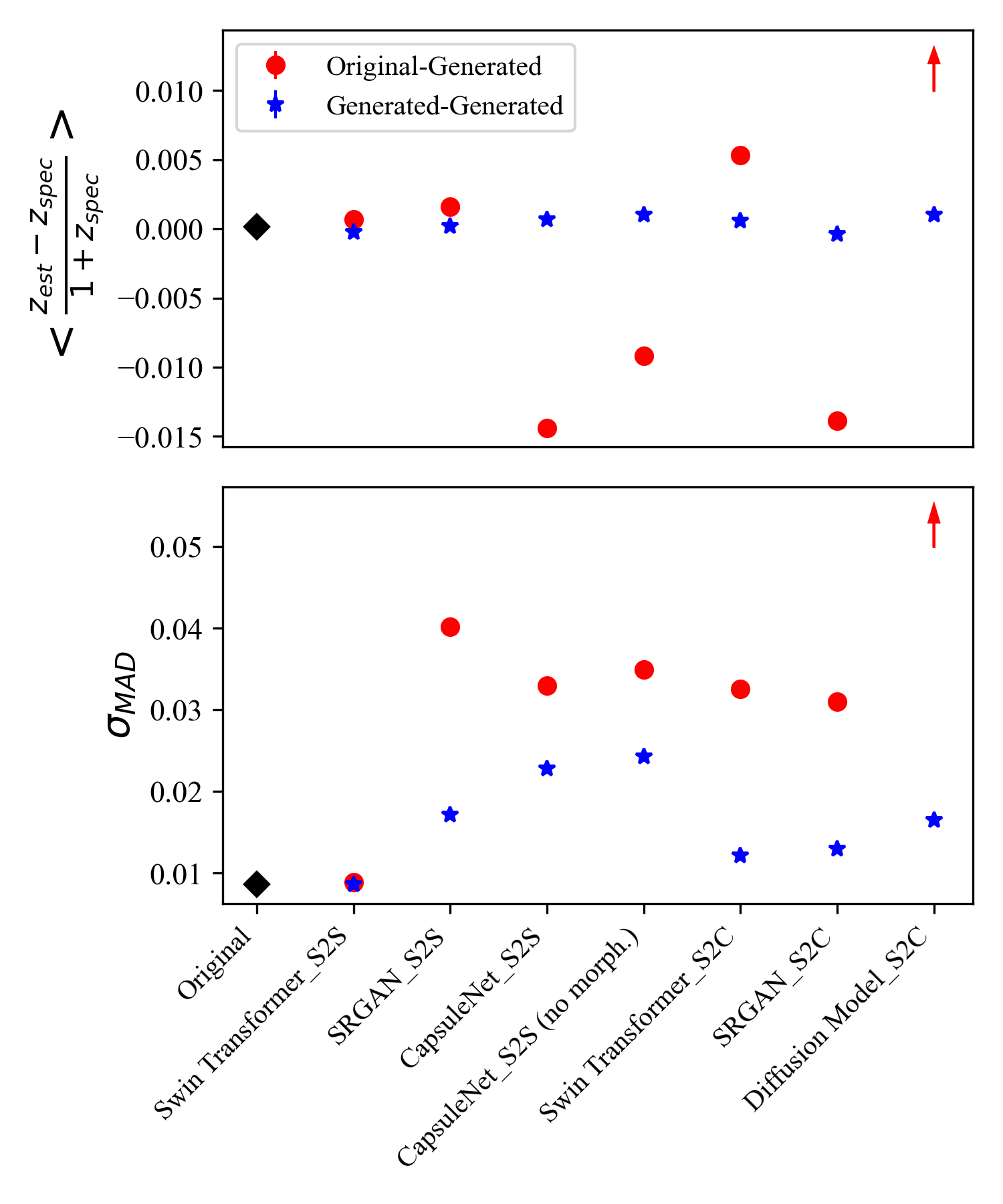}
\caption{Global mean redshift residuals and $\sigma_{\mathrm{MAD}}$ (defined in the main text) for all the image translation models. The results from the capsule network implementation without using the morphological class labels are also shown, denoted as ``CapsuleNet$\_$S2S (no morph.)''. The redshift predictions were made for the generated test images using redshift estimation models trained with either the original or the generated $z$-training images, denoted as ``original-generated'' and ``generated-generated'', respectively. These results are compared with the predictions for the original SDSS test images by a model trained with the original SDSS $z$-training images (i.e. the baseline). The errorbars are too small to be visible. The mean residual and $\sigma_{\mathrm{MAD}}$ from the diffusion model in the original-generated case are dramatically beyond the plotted ranges (i.e. 0.193 and 0.214, respectively) and indicated by the red arrows.}
\label{fig:error_z_sigma}
\end{figure}

\begin{figure*}
\begin{center}
\centerline{\includegraphics[width=1.0\linewidth]{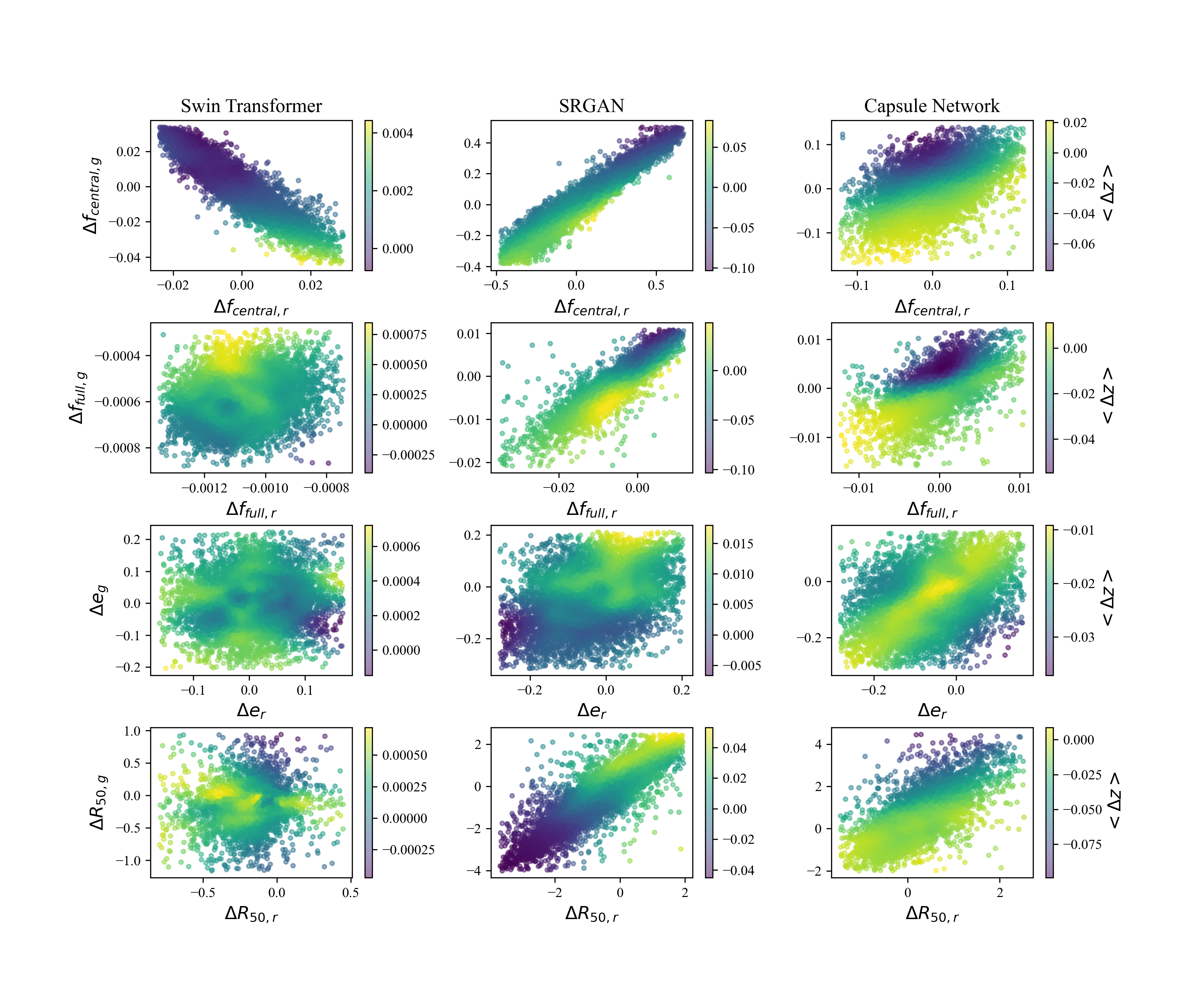}}
\caption{Analysis of redshift dependences on flux and galaxy property residuals using the S2S translated $g$-band and $r$-band images and original SDSS validation images. \textit{First row:} Mean central flux residuals between the generated and original validation images in $g$ band as a function of the residuals in $r$ band, shown for the Swin Transformer, the SRGAN, and the capsule network. The fluxes (rescaled with Eq.~\ref{eq:norm}) are shown in the natural logarithmic scale. All the data points are color coded with the local mean difference between the redshift predictions for the generated and original SDSS validation images by a redshift estimation model trained with the original SDSS $z$-training images. The colors are smoothed using the \texttt{LOESS} method from \citet{Cappellari2013}. \textit{Second row:} Same as the first row, but shown for the mean fluxes over the full images. \textit{Third row:} Same as the first row, but shown for the ellipticity $e$. \textit{Fourth row:} Same as the first row, but shown for the half-light radius $R_{50}$.}
\label{fig:diff_z_s2s}
\end{center}
\end{figure*}

\begin{figure*}
\begin{center}
\centerline{\includegraphics[width=1.0\linewidth]{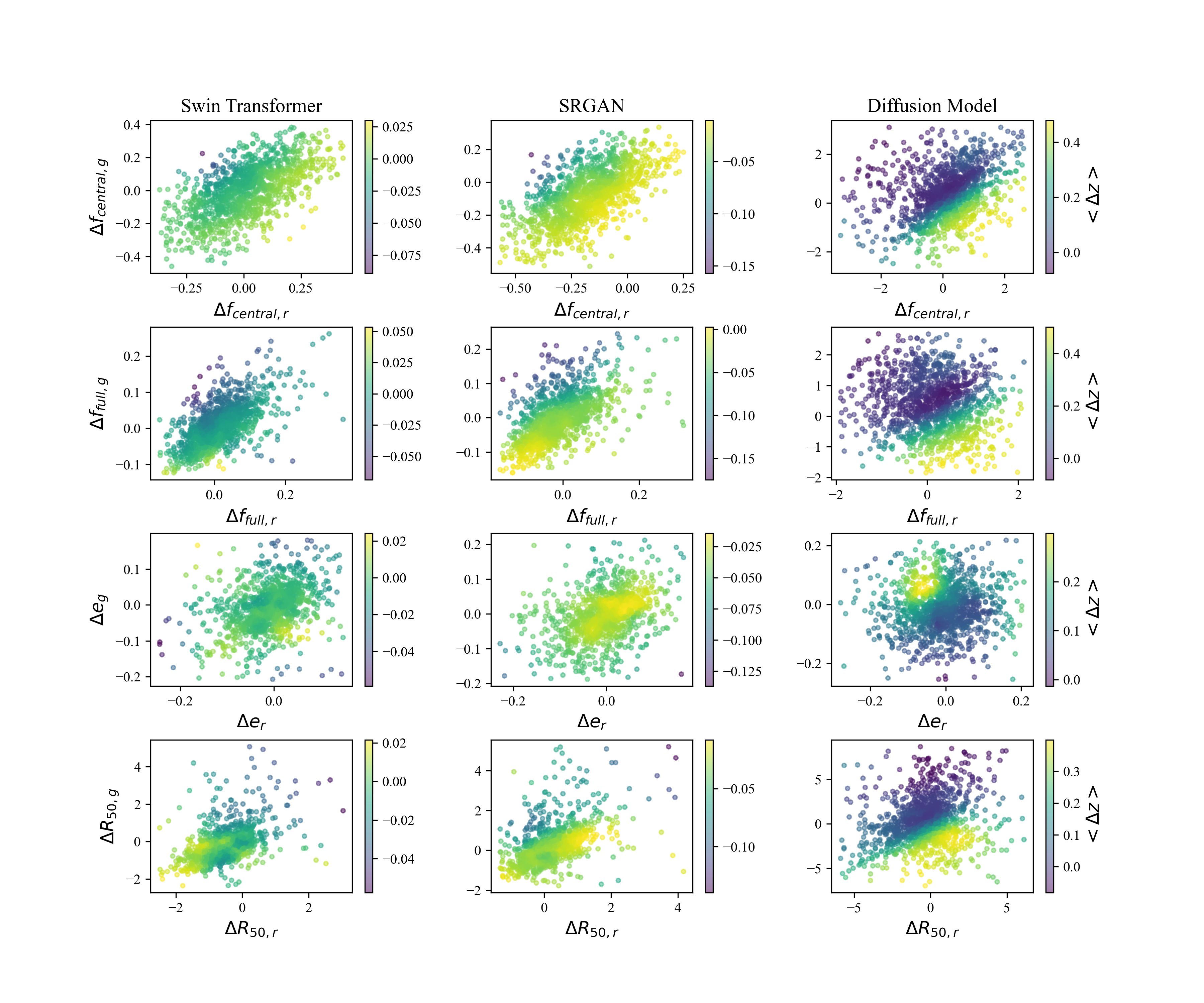}}
\caption{Same as Fig.~\ref{fig:diff_z_s2s}, but shown for the S2C translated images and original CFHTLS validation images. All the data points are color coded with the local mean difference between the redshift predictions for the generated and original CFHTLS validation images by a redshift estimation model trained with the original CFHTLS $z$-training images.}
\label{fig:diff_z_s2c}
\end{center}
\end{figure*}

\begin{figure*}
\begin{center}
\centerline{\includegraphics[width=1.0\linewidth]{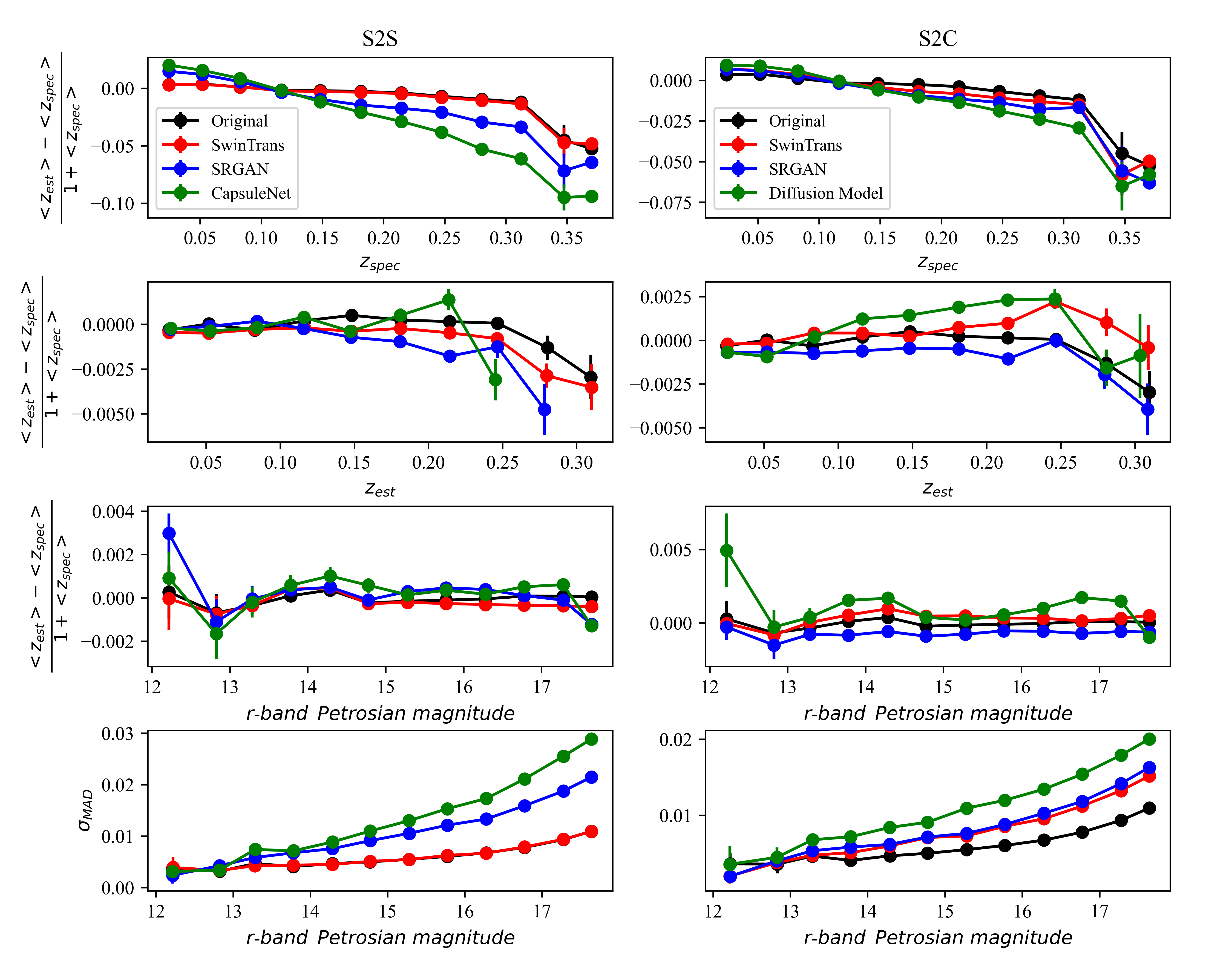}}
\caption{Bin-wise mean redshift residuals and $\sigma_{\mathrm{MAD}}$ (defined in the main text) for all the image translation models. For each translation model, the redshift predictions were made for the generated test images using redshift estimation models trained with the generated $z$-training images. A comparison is made with the predictions for the original SDSS test images by a model trained with the original SDSS $z$-training images (i.e. the baseline). \textit{First row:} Bin-wise mean redshift residuals as a function of spectroscopic redshift $z_{spec}$ for the S2S and S2C translations, separately. The error bars are estimated using the root mean square error of residuals in each redshift bin. \textit{Second row:} Bin-wise mean redshift residuals as a function of the predicted redshift point estimate $z_{est}$. \textit{Third row:} Bin-wise mean redshift residuals as a function of $r$-band Petrosian magnitude. \textit{Fourth row:} MAD-based dispersion, $\sigma_{\mathrm{MAD}}$, as a function of $r$-band Petrosian magnitude. The curve obtained by the Swin Transformer in the S2S translation almost overlaps with the baseline.}
\label{fig:ori_re}
\end{center}
\end{figure*}

To assess the preservation of redshift information, we analyzed the redshift predictions for the generated images from the test sample using redshift estimation models trained with either the original or the generated $z$-training images (i.e. original-generated and generated-generated, respectively). We quantified redshift information using the redshift residuals and the median absolute deviation (MAD)-based dispersion, defined as
\begin{equation}
\delta z = \frac{z_{est} - z_{spec}}{1 + z_{spec}},
\label{eq:instancewise_residual}
\end{equation}
and
\begin{equation}
\sigma_{\mathrm{MAD}} = 1.4826 \times \mathrm{Median} \left| \delta z - \mathrm{Median}(\delta z) \right|,
\label{eq:sigma_mad}
\end{equation}
where $z_{est}$ stands for the redshift point estimate; $z_{spec}$ stands for the reference spectroscopic redshift. The global mean redshift residuals and $\sigma_{\mathrm{MAD}}$ for all the image translation models are illustrated in Fig.~\ref{fig:error_z_sigma}, compared with a baseline in which both the training and predictions only involved the original SDSS images. The results from the capsule network implementations with and without using the morphological classifications are both shown.

In the original-generated case, the Swin Transformer in the S2S translation has an excellent performance in reproducing the features that encode redshift information, reaching a mean residual close to zero and a $\sigma_{\mathrm{MAD}}$ that is only slightly higher than that for the original images (i.e. the baseline). This is due to the simplicity of such a scenario in which the domain to be translated is identical to the target domain. In contrast, the $\sigma_{\mathrm{MAD}}$ is much higher for the SRGAN and the capsule network, implying that these two models have modified the image features that encode redshift information. The generative adversarial training process of the SRGAN does not specify redshift information and may make it overconcentrate on unphysical features. The capsule network may undergo a heavy compression of information. As revealed by the comparison between the two capsule network implementations, the use of morphological class labels does not lead to a noticeable improvement on the preservation of redshift information. In the S2C translation, all the models produce high $\sigma_{\mathrm{MAD}}$ and the mean residuals are not well constrained, as a result of the diversity between the SDSS and CFHTLS image domains and a limited number of paired images for training. In particular, the diffusion model has the worst performance due to its instability in calibrating the fluxes.

However, the $\sigma_{\mathrm{MAD}}$ for all the models can be significantly improved and a good constraint on the mean residuals can be restored when redshift estimation models are trained with the generated images rather than the original images, as shown in the generated-generated case. It is interesting to notice that the models in the S2C translation including the diffusion model obtain even lower $\sigma_{\mathrm{MAD}}$ than the SRGAN and the capsule network do in the S2S translation. This suggests that the S2C translated images, though having more uncertainties and different from the original ones, still contain a considerable amount of redshift information. Nonetheless, the gap in $\sigma_{\mathrm{MAD}}$ between all the models in the generated-generated case and the baseline indicates a loss of redshift information in image translation.

\subsubsection{Analysis of the original-generated case} \label{sec:ori_gen}

To conduct further analysis, we took the $g$-band and $r$-band images translated via S2S and S2C from the original SDSS validation images, and used them as examples to demonstrate redshift dependences on the mean central fluxes, the mean full-image fluxes, the ellipticity $e$, and the half-light radius $R_{50}$. The results are illustrated in Figs.~\ref{fig:diff_z_s2s} and \ref{fig:diff_z_s2c}. For each of these properties, the residuals between the generated and original images in $g$ band are contrasted with the residuals in $r$ band, color coded with the local mean difference between the redshifts predicted for the generated and original images. We applied the \texttt{LOESS} method\footnote{https://pypi.org/project/loess} \citep{Cappellari2013} to smooth the colors.

In the S2S translation, the redshift differences generally show intricate connections with the two-band flux residuals. In particular, for the Swin Transformer, the dependence of redshift on the mean full-image fluxes is not conspicuous, whereas positive redshift differences tend to appear with negative $g$-band central flux residuals $\Delta f_{central,g}$ and positive $r$-band residuals $\Delta f_{central,r}$. In other words, redshifts tend to be overestimated when the reproduced $g-r$ colors are slightly higher than the original values. The two-band mean central flux residuals combined reveal the tiny deviations between the S2S translated images by the Swin Transformer and the original SDSS images that impact redshift predictions, suggesting that the cross-band peak fluxes contain meaningful redshift information. For the SRGAN and the capsule network, negative redshift differences tend to appear with positive $\Delta f_{central,g}$, but have different dependences on $\Delta f_{central,r}$. Similar behaviors can be seen with the full-image flux residuals $\Delta f_{full,g}$ and $\Delta f_{full,r}$ for these two models. For the morphological properties, the redshift differences show stronger dependences on the half-light radius $R_{50}$ than on the ellipticity $e$, which may also be correlated with the flux residuals. Namely, for the SRGAN, negative redshift differences tend to appear with negative $\Delta R_{50,g}$ and $\Delta R_{50,r}$ that are associated with positive $\Delta f_{central,g}$ and $\Delta f_{central,r}$; while for the capsule network, negative redshift differences tend to appear with positive $\Delta R_{50,g}$ that is roughly associated with positive $\Delta f_{central,g}$. That is, redshifts tend to be underestimated when the SRGAN uplifts the $g$-band and $r$-band peak fluxes (which lowers the effective galaxy sizes), or when the capsule network enlarges both the galaxy sizes and brightness in $g$ band.

In the S2C translation, it is harder to identify redshift dependences using the aforementioned two-band residual plots due to large uncertainties. Nonetheless, it is recognizable, for example, that increasing redshift differences seem to be associated with decreasing $\Delta f_{central,g}$ and increasing $\Delta f_{central,r}$, as well as decreasing $\Delta R_{50,g}$, again pointing to the connections between redshift predictions and $g-r$ colors.

\subsubsection{Analysis of the generated-generated case} \label{sec:gen_gen}

Finally, we analyzed the behaviors of the bin-wise mean redshift residuals and $\sigma_{\mathrm{MAD}}$ in the generated-generated case using the SDSS test sample, in order to examine how the redshift information preserved by the image translation models (regardless of its manifestation) is distributed across the sample. The results are illustrated in Fig.~\ref{fig:ori_re}. The bin-wise mean redshift residuals are defined as
\begin{equation}
\delta_{<z>} = \frac{<z_{est}> - <z_{spec}>}{1 + <z_{spec}>},
\label{eq:binwise_residual}
\end{equation}
where $<>$ denotes the mean computed in a redshift or magnitude bin. This definition is different from Eq.~\ref{eq:instancewise_residual}.

We found that the bin-wise mean redshift residuals for all the models decrease monotonically as a function of spectroscopic redshift $z_{spec}$, and remain nearly flat as a function of the predicted redshift point estimate $z_{est}$ or $r$-band Petrosian magnitude. These trends are in resemblance to the baseline in which only the original SDSS images were involved in training and redshift predictions. The gaps in the $\sigma_{\mathrm{MAD}}$ curves between any different models gradually increase as a function of $r$-band Petrosian magnitude, associated with the decreasing slopes of the residual-$z_{spec}$ curves. Importantly, this indicates that the information loss experienced by all the models is monotonic along galaxy brightness and $z_{spec}$, exhibiting no apparent local information imbalance that significantly biases redshift predictions.

In summary, the image translation process may alter the manifestation of redshift information on galaxy images and result in information loss (e.g. by miscalibrating the peak fluxes), and thus downstream models (e.g. redshift estimation) trained with original images may not be directly applicable to translated images. However, redshift information may be present in the form of other image features (e.g. global flux profile shapes), and thus translated images can still be leveraged for downstream tasks that do not strongly require high image fidelity, as long as sufficient information is preserved and information loss does not bias the downstream tasks. In this sense, a model with good efficiency to learn and retain information is essential for image translation.

\section{Conclusion} \label{sec:conclusion}

In this work, we investigated four representative generative models for translating galaxy images, i.e. a Swin Transformer, an SRGAN, a capsule network, and a diffusion model, performing an intra-domain SDSS-to-SDSS translation (S2S) and an inter-domain SDSS-to-CFHTLS translation (S2C). By comparing the generated and original images in the aspects of pixel-level, morphology-level, and high-order physical information, we obtained the results as follows.

\begin{itemize}
\item In general, the Swin Transformer outperforms other models considering all the results, probably because this model has the best flexibility and stability. In contrast, the SRGAN has some level of instability; the generative adversarial training process does not specify redshift information for the SRGAN but may instead make it overconcentrate on unphysical features, overly learn the properties of over-represented galaxies, and aggravate attenuation biases. The capsule network has a heavy loss of information, and the use of morphological classifications has no noticeable contribution to the preservation of redshift information. Even though the diffusion model can generate noise patterns that resemble the original images, the shortcoming of this model is a lack of stability that is unfavorable for the calibration of fluxes and the preservation of redshift information. We do not intend to mean that a transformer is universally superior over other neural networks for image translation, but emphasize that the flexibility and stability of a model are crucial for generating images with good fidelity.

\item At the pixel level, all the image translation models especially the Swin Transformer can generate images that are generally comparable with the original ones. However, by analyzing the mean fluxes estimated over the full image and over the four central pixels, we found that it is more difficult to constrain the peak fluxes than the global structures of galaxies. Furthermore, it is not trivial to bypass errors and biases in calibrating the fluxes especially in the S2C translation. This is because image translation has the nature of many-to-many mapping and the SDSS-CFHTLS paired training data are limited. The problems at the pixel level are related to those associated with morphology and redshift information.
 
\item We analyzed the semimajor axis $a$, the ellipticity $e$, the AB magnitude based on the Kron flux, the FWHM, and the half-light radius $R_{50}$, and found that all the models especially the Swin Transformer can approximately retain the morphology-level statistics of the original data. Nonetheless, for any of the five properties, all the models cannot avoid attenuation biases and dispersions between the generated and original images, which may be due to measurement uncertainties for both the images. This again links to the many-to-many problem.

\item Regarding spectroscopic redshift as a representative of high-order physical information, we used redshift estimation models trained with either original or generated images to predict redshifts for the images generated by the image translation models (i.e. original-generated and generated-generated). These two cases were used to test the ability of a model to reproduce the features that encode redshift information in comparison with the original images, and the ability to preserve redshift information regardless of how it is manifested on the generated images. In the original-generated case, the Swin Transformer performs the best in the S2S translation, producing a mean redshift residual close to zero and a $\sigma_{\mathrm{MAD}}$ almost the same as the baseline (i.e. obtained using the original images). On the contrary, all the models perform relatively poorly in the S2C translation. We analyzed redshift dependences on the mean central fluxes, the mean full-image fluxes, the ellipticity $e$, and the half-light radius $R_{50}$ in $g$ and $r$ bands. In particular, we found intricate trends between redshift and cross-band central fluxes, implying that the central fluxes contain meaningful redshift information.

\item In the generated-generated case, the mean redshift residual and $\sigma_{\mathrm{MAD}}$ can be improved for all the models. Interestingly, the models in the S2C translation achieve even better preservation of redshift information than the models in the S2S translation except the Swin Transformer, suggesting that the S2C translation models can convey a considerable amount of redshift information despite alterations of the original image features and a loss of information. Furthermore, the bin-wise mean redshift residuals and $\sigma_{MAD}$ for all the models show monotonic dependences on galaxy brightness and spectroscopic redshift, implying that the loss of redshift information is monotonous and there is no apparent local information imbalance that causes considerable biases in redshift predictions.

\end{itemize}

To conclude, this work provides meaningful insights into the manifestation of complex high-order physical information on galaxy images and points out the inadequacies of current image translation models. In particular, we point out that a good calibration of cross-band peak fluxes of galaxies is crucial for retaining redshift information and ensuring the fidelity of generated images, but it is subject to noticeable uncertainties. Nonetheless, translated images, despite being not realistic, may still contain meaningful physical information. We therefore suggest that these imperfect images may still be exploited for downstream tasks that do not have strong requirements of high image fidelity (e.g. via data augmentation and missing data imputation) on condition that no significant bias is introduced. To this end, preserving sufficient information is a prerequisite. Thus for optimizing image translation models, we emphasize that more care should be taken to ensure the flexibility and stability of a model so that it has the capability to learn and retain sufficient information. 

Despite these considerations, in order to promote data-driven research and scientific discoveries that require high-fidelity data enhancement or information extraction, it is of great significance to preserve complex high-order physical information in image translation. We will conduct further investigations on the preservation of physical information in future work.

\begin{acknowledgements}

This work is supported by ``The Major Key Project of PCL''.

R.H.X. acknowledges financial support from the National Natural Science Foundation of China (62201306).

This research makes use of Photutils, an Astropy package for detection and photometry of astronomical sources \citep{larry_bradley_2024_13989456}.

This work makes use of the Sloan Digital Sky Survey (SDSS) data. Funding for SDSS-III has been provided by the Alfred P. Sloan Foundation, the Participating Institutions, the National Science Foundation, and the U.S. Department of Energy Office of Science. The SDSS-III web site is http://www.sdss3.org/. SDSS-III is managed by the Astrophysical Research Consortium for the Participating Institutions of the SDSS-III Collaboration including the University of Arizona, the Brazilian Participation Group, Brookhaven National Laboratory, Carnegie Mellon University, University of Florida, the French Participation Group, the German Participation Group, Harvard University, the Instituto de Astrofisica de Canarias, the Michigan State/Notre Dame/JINA Participation Group, Johns Hopkins University, Lawrence Berkeley National Laboratory, Max Planck Institute for Astrophysics, Max Planck Institute for Extraterrestrial Physics, New Mexico State University, New York University, Ohio State University, Pennsylvania State University, University of Portsmouth, Princeton University, the Spanish Participation Group, University of Tokyo, University of Utah, Vanderbilt University, University of Virginia, University of Washington, and Yale University.

This work is based on observations obtained with MegaPrime/MegaCam, a joint project of CFHT and CEA/DAPNIA, at the Canada-France-Hawaii Telescope (CFHT) which is operated by the National Research Council (NRC) of Canada, the Institute National des Sciences de l'Univers of the Centre National de la Recherche Scientifique (CNRS) of France, and the University of Hawaii. This work is based in part on data products produced at Terapix and the Canadian Astronomy Data Centre as part of the Canada-France-Hawaii Telescope Legacy Survey, a collaborative project of NRC and CNRS.

\end{acknowledgements}

\bibliographystyle{aa}
\bibliography{aa54947-25.bib}

\begin{thebibliography}{82}
\expandafter\ifx\csname natexlab\endcsname\relax\def\natexlab#1{#1}\fi

\bibitem[{{Adam} {et~al.}(2025){Adam}, {Stone}, {Bottrell}, {Legin}, {Hezaveh},
  \& {Perreaul-Levasseur}}]{Adam2025}
{Adam}, A., {Stone}, C., {Bottrell}, C., {et~al.} 2025, \aj, 169, 254

\bibitem[{{Akhaury} {et~al.}(2024){Akhaury}, {Jablonka}, {Starck}, \&
  {Courbin}}]{Akhaury2024}
{Akhaury}, U., {Jablonka}, P., {Starck}, J.~L., \& {Courbin}, F. 2024, \aap,
  688, A6

\bibitem[{{Akhaury} {et~al.}(2022){Akhaury}, {Starck}, {Jablonka}, {Courbin},
  \& {Michalewicz}}]{Akhaury2022}
{Akhaury}, U., {Starck}, J.-L., {Jablonka}, P., {Courbin}, F., \&
  {Michalewicz}, K. 2022, Frontiers in Astronomy and Space Sciences, 9, 357

\bibitem[{{Alam} {et~al.}(2015){Alam}, {Albareti}, {Allende Prieto}, {Anders},
  {Anderson}, {Anderton}, {Andrews}, {Armengaud}, {Aubourg}, {Bailey}, {Basu},
  {Bautista}, {Beaton}, {Beers}, {Bender}, {Berlind}, {Beutler}, {Bhardwaj},
  {Bird}, {Bizyaev}, {Blake}, {Blanton}, {Blomqvist}, {Bochanski}, {Bolton},
  {Bovy}, {Shelden Bradley}, {Brandt}, {Brauer}, {Brinkmann}, {Brown},
  {Brownstein}, {Burden}, {Burtin}, {Busca}, {Cai}, {Capozzi}, {Carnero
  Rosell}, {Carr}, {Carrera}, {Chambers}, {Chaplin}, {Chen}, {Chiappini},
  {Chojnowski}, {Chuang}, {Clerc}, {Comparat}, {Covey}, {Croft}, {Cuesta},
  {Cunha}, {da Costa}, {Da Rio}, {Davenport}, {Dawson}, {De Lee}, {Delubac},
  {Deshpande}, {Dhital}, {Dutra-Ferreira}, {Dwelly}, {Ealet}, {Ebelke},
  {Edmondson}, {Eisenstein}, {Ellsworth}, {Elsworth}, {Epstein}, {Eracleous},
  {Escoffier}, {Esposito}, {Evans}, {Fan}, {Fern{\'a}ndez-Alvar}, {Feuillet},
  {Filiz Ak}, {Finley}, {Finoguenov}, {Flaherty}, {Fleming}, {Font-Ribera},
  {Foster}, {Frinchaboy}, {Galbraith-Frew}, {Garc{\'\i}a},
  {Garc{\'\i}a-Hern{\'a}ndez}, {Garc{\'\i}a P{\'e}rez}, {Gaulme}, {Ge},
  {G{\'e}nova-Santos}, {Georgakakis}, {Ghezzi}, {Gillespie}, {Girardi},
  {Goddard}, {Gontcho}, {Gonz{\'a}lez Hern{\'a}ndez}, {Grebel}, {Green},
  {Grieb}, {Grieves}, {Gunn}, {Guo}, {Harding}, {Hasselquist}, {Hawley},
  {Hayden}, {Hearty}, {Hekker}, {Ho}, {Hogg}, {Holley-Bockelmann}, {Holtzman},
  {Honscheid}, {Huber}, {Huehnerhoff}, {Ivans}, {Jiang}, {Johnson},
  {Kinemuchi}, {Kirkby}, {Kitaura}, {Klaene}, {Knapp}, {Kneib}, {Koenig},
  {Lam}, {Lan}, {Lang}, {Laurent}, {Le Goff}, {Leauthaud}, {Lee}, {Lee},
  {Licquia}, {Liu}, {Long}, {L{\'o}pez-Corredoira}, {Lorenzo-Oliveira},
  {Lucatello}, {Lundgren}, {Lupton}, {Mack}, {Mahadevan}, {Maia}, {Majewski},
  {Malanushenko}, {Malanushenko}, {Manchado}, {Manera}, {Mao}, {Maraston},
  {Marchwinski}, {Margala}, {Martell}, {Martig}, {Masters}, {Mathur},
  {McBride}, {McGehee}, {McGreer}, {McMahon}, {M{\'e}nard}, {Menzel},
  {Merloni}, {M{\'e}sz{\'a}ros}, {Miller}, {Miralda-Escud{\'e}}, {Miyatake},
  {Montero-Dorta}, {More}, {Morganson}, {Morice-Atkinson}, {Morrison},
  {Mosser}, {Muna}, {Myers}, {Nandra}, {Newman}, {Neyrinck}, {Nguyen},
  {Nichol}, {Nidever}, {Noterdaeme}, {Nuza}, {O'Connell}, {O'Connell},
  {O'Connell}, {Ogando}, {Olmstead}, {Oravetz}, {Oravetz}, {Osumi}, {Owen},
  {Padgett}, {Padmanabhan}, {Paegert}, {Palanque-Delabrouille}, {Pan},
  {Parejko}, {P{\^a}ris}, {Park}, {Pattarakijwanich}, {Pellejero-Ibanez},
  {Pepper}, {Percival}, {P{\'e}rez-Fournon}, {P{\'e}rez-R{\`a}fols},
  {Petitjean}, {Pieri}, {Pinsonneault}, {Porto de Mello}, {Prada}, {Prakash},
  {Price-Whelan}, {Protopapas}, {Raddick}, {Rahman}, {Reid}, {Rich}, {Rix},
  {Robin}, {Rockosi}, {Rodrigues}, {Rodr{\'\i}guez-Torres}, {Roe}, {Ross},
  {Ross}, {Rossi}, {Ruan}, {Rubi{\~n}o-Mart{\'\i}n}, {Rykoff},
  {Salazar-Albornoz}, {Salvato}, {Samushia}, {S{\'a}nchez}, {Santiago},
  {Sayres}, {Schiavon}, {Schlegel}, {Schmidt}, {Schneider}, {Schultheis},
  {Schwope}, {Sc{\'o}ccola}, {Scott}, {Sellgren}, {Seo}, {Serenelli}, {Shane},
  {Shen}, {Shetrone}, {Shu}, {Silva Aguirre}, {Sivarani}, {Skrutskie},
  {Slosar}, {Smith}, {Sobreira}, {Souto}, {Stassun}, {Steinmetz}, {Stello},
  {Strauss}, {Streblyanska}, {Suzuki}, {Swanson}, {Tan}, {Tayar}, {Terrien},
  {Thakar}, {Thomas}, {Thomas}, {Thompson}, {Tinker}, {Tojeiro}, {Troup},
  {Vargas-Maga{\~n}a}, {Vazquez}, {Verde}, {Viel}, {Vogt}, {Wake}, {Wang},
  {Weaver}, {Weinberg}, {Weiner}, {White}, {Wilson}, {Wisniewski},
  {Wood-Vasey}, {Ye`che}, {York}, {Zakamska}, {Zamora}, {Zasowski}, {Zehavi},
  {Zhao}, {Zheng}, {Zhou}, {Zhou}, {Zou}, \& {Zhu}}]{Alam2015}
{Alam}, S., {Albareti}, F.~D., {Allende Prieto}, C., {et~al.} 2015, \apjs, 219,
  12

\bibitem[{{Alfonzo} {et~al.}(2024){Alfonzo}, {Iyer}, {Akiyama}, {Bryan},
  {Cooray}, {Ludwig}, {Mowla}, {Omori}, {Pacifici}, {Speagle}, \&
  {Wu}}]{Alfonzo2024}
{Alfonzo}, J.~P., {Iyer}, K.~G., {Akiyama}, M., {et~al.} 2024, \apj, 967, 152

\bibitem[{{Arcelin} {et~al.}(2021){Arcelin}, {Doux}, {Aubourg}, {Roucelle}, \&
  {LSST Dark Energy Science Collaboration}}]{Arcelin2021}
{Arcelin}, B., {Doux}, C., {Aubourg}, E., {Roucelle}, C., \& {LSST Dark Energy
  Science Collaboration}. 2021, \mnras, 500, 531

\bibitem[{{Boucaud} {et~al.}(2020){Boucaud}, {Huertas-Company}, {Heneka},
  {Ishida}, {Sedaghat}, {de Souza}, {Moews}, {Dole}, {Castellano}, {Merlin},
  {Roscani}, {Tramacere}, {Killedar}, {Trindade}, \& {Collaboration
  COIN}}]{Boucaud2020}
{Boucaud}, A., {Huertas-Company}, M., {Heneka}, C., {et~al.} 2020, \mnras, 491,
  2481

\bibitem[{Bradley {et~al.}(2024)Bradley, Sip{\H o}cz, Robitaille, Tollerud,
  Vin{\'{\i}}cius, Deil, Barbary, Wilson, Busko, Donath, G{\"u}nther, Cara,
  Lim, Me{\ss}linger, Conseil, Burnett, Bostroem, Droettboom, Bray, Bratholm,
  Ginsburg, Jamieson, Barentsen, Craig, Morris, Perrin, Rathi, Pascual, \&
  Georgiev}]{larry_bradley_2024_13989456}
Bradley, L., Sip{\H o}cz, B., Robitaille, T., {et~al.} 2024, astropy/photutils:
  2.0.2

\bibitem[{{Buncher} {et~al.}(2021){Buncher}, {Sharma}, \& {Carrasco
  Kind}}]{Buncher2021}
{Buncher}, B., {Sharma}, A.~N., \& {Carrasco Kind}, M. 2021, \mnras, 503, 777

\bibitem[{{Campagne}(2020)}]{Campagne2020}
{Campagne}, J.-E. 2020, arXiv e-prints, arXiv:2002.10154

\bibitem[{{Cappellari} {et~al.}(2013){Cappellari}, {McDermid}, {Alatalo},
  {Blitz}, {Bois}, {Bournaud}, {Bureau}, {Crocker}, {Davies}, {Davis}, {de
  Zeeuw}, {Duc}, {Emsellem}, {Khochfar}, {Krajnovi{\'c}}, {Kuntschner},
  {Morganti}, {Naab}, {Oosterloo}, {Sarzi}, {Scott}, {Serra}, {Weijmans}, \&
  {Young}}]{Cappellari2013}
{Cappellari}, M., {McDermid}, R.~M., {Alatalo}, K., {et~al.} 2013, \mnras, 432,
  1862

\bibitem[{{Cheng} {et~al.}(2021){Cheng}, {Huertas-Company}, {Conselice},
  {Arag{\'o}n-Salamanca}, {Robertson}, \& {Ramachandra}}]{Cheng2021}
{Cheng}, T.-Y., {Huertas-Company}, M., {Conselice}, C.~J., {et~al.} 2021,
  \mnras, 503, 4446

\bibitem[{{{\'C}iprijanovi{\'c}} {et~al.}(2022){{\'C}iprijanovi{\'c}},
  {Kafkes}, {Snyder}, {S{\'a}nchez}, {Perdue}, {Pedro}, {Nord}, {Madireddy}, \&
  {Wild}}]{Aleksandra2022}
{{\'C}iprijanovi{\'c}}, A., {Kafkes}, D., {Snyder}, G., {et~al.} 2022, Machine
  Learning: Science and Technology, 3, 035007

\bibitem[{{D'Addona} {et~al.}(2021){D'Addona}, {Riccio}, {Cavuoti}, {Tortora},
  \& {Brescia}}]{DAddona2021}
{D'Addona}, M., {Riccio}, G., {Cavuoti}, S., {Tortora}, C., \& {Brescia}, M.
  2021, in Intelligent Astrophysics, ed. I.~{Zelinka}, M.~{Brescia}, \&
  D.~{Baron}, Vol.~39, 225--244

\bibitem[{{Dey} {et~al.}(2022){Dey}, {Andrews}, {Newman}, {Mao}, {Rau}, \&
  {Zhou}}]{Dey2022}
{Dey}, B., {Andrews}, B.~H., {Newman}, J.~A., {et~al.} 2022, \mnras, 515, 5285

\bibitem[{{Dia} {et~al.}(2020){Dia}, {Savary}, {Melchior}, \&
  {Courbin}}]{Dia2020}
{Dia}, M., {Savary}, E., {Melchior}, M., \& {Courbin}, F. 2020, in Astronomical
  Society of the Pacific Conference Series, Vol. 527, Astronomical Data
  Analysis Software and Systems XXIX, ed. R.~{Pizzo}, E.~R. {Deul}, J.~D.
  {Mol}, J.~{de Plaa}, \& H.~{Verkouter}, 175

\bibitem[{{Fang} {et~al.}(2023){Fang}, {Ba}, {Gu}, {Lin}, {Hou}, {Qin}, {Zhou},
  {Xu}, {Dai}, {Song}, \& {Kong}}]{Fang2023}
{Fang}, G., {Ba}, S., {Gu}, Y., {et~al.} 2023, \aj, 165, 35

\bibitem[{{Fussell} \& {Moews}(2019)}]{Fussell2019}
{Fussell}, L. \& {Moews}, B. 2019, \mnras, 485, 3203

\bibitem[{{Gan} {et~al.}(2021){Gan}, {Bekki}, \& {Hashemizadeh}}]{Gan2021}
{Gan}, F.~K., {Bekki}, K., \& {Hashemizadeh}, A. 2021, arXiv e-prints,
  arXiv:2103.09711

\bibitem[{{Germain} {et~al.}(2015){Germain}, {Gregor}, {Murray}, \&
  {Larochelle}}]{Germain2015}
{Germain}, M., {Gregor}, K., {Murray}, I., \& {Larochelle}, H. 2015, arXiv
  e-prints, arXiv:1502.03509

\bibitem[{Goodfellow {et~al.}(2014)Goodfellow, Pouget-Abadie, Mirza, Xu,
  Warde-Farley, Ozair, Courville, \& Bengio}]{Goodfellow2014}
Goodfellow, I.~J., Pouget-Abadie, J., Mirza, M., {et~al.} 2014, in Proceedings
  of the 28th International Conference on Neural Information Processing Systems
  - Volume 2, NIPS'14 (Cambridge, MA, USA: MIT Press), 2672–2680

\bibitem[{{Graff} {et~al.}(2014){Graff}, {Feroz}, {Hobson}, \&
  {Lasenby}}]{Graff2014}
{Graff}, P., {Feroz}, F., {Hobson}, M.~P., \& {Lasenby}, A. 2014, \mnras, 441,
  1741

\bibitem[{{Gwyn}(2012)}]{Gwyn2012}
{Gwyn}, S. D.~J. 2012, \aj, 143, 38

\bibitem[{Heusel {et~al.}(2017)Heusel, Ramsauer, Unterthiner, Nessler, \&
  Hochreiter}]{Heusel2017}
Heusel, M., Ramsauer, H., Unterthiner, T., Nessler, B., \& Hochreiter, S. 2017,
  in Advances in Neural Information Processing Systems, ed. I.~Guyon, U.~V.
  Luxburg, S.~Bengio, H.~Wallach, R.~Fergus, S.~Vishwanathan, \& R.~Garnett,
  Vol.~30 (Curran Associates, Inc.)

\bibitem[{Ho {et~al.}(2020)Ho, Jain, \& Abbeel}]{Ho2020}
Ho, J., Jain, A., \& Abbeel, P. 2020, in Proceedings of the 34th International
  Conference on Neural Information Processing Systems, NIPS '20 (Red Hook, NY,
  USA: Curran Associates Inc.)

\bibitem[{{Holzschuh} {et~al.}(2022){Holzschuh}, {O'Riordan}, {Vegetti},
  {Rodriguez-Gomez}, \& {Thuerey}}]{Holzschuh2022}
{Holzschuh}, B.~J., {O'Riordan}, C.~M., {Vegetti}, S., {Rodriguez-Gomez}, V.,
  \& {Thuerey}, N. 2022, \mnras, 515, 652

\bibitem[{{Hudelot} {et~al.}(2012){Hudelot}, {Cuillandre}, {Withington},
  {Goranova}, {McCracken}, {Magnard}, {Mellier}, {Regnault}, {Betoule},
  {Aussel}, {Kavelaars}, {Fernique}, {Bonnarel}, {Ochsenbein}, \&
  {Ilbert}}]{CFHTLST07}
{Hudelot}, P., {Cuillandre}, J.~C., {Withington}, K., {et~al.} 2012, VizieR
  Online Data Catalog, II/317

\bibitem[{{Ivezi{\'c}} {et~al.}(2019){Ivezi{\'c}}, {Kahn}, {Tyson}, {Abel},
  {Acosta}, {Allsman}, {Alonso}, {AlSayyad}, {Anderson}, {Andrew}, {Angel},
  {Angeli}, {Ansari}, {Antilogus}, {Araujo}, {Armstrong}, {Arndt}, {Astier},
  {Aubourg}, {Auza}, {Axelrod}, {Bard}, {Barr}, {Barrau}, {Bartlett}, {Bauer},
  {Bauman}, {Baumont}, {Bechtol}, {Bechtol}, {Becker}, {Becla}, {Beldica},
  {Bellavia}, {Bianco}, {Biswas}, {Blanc}, {Blazek}, {Blandford}, {Bloom},
  {Bogart}, {Bond}, {Booth}, {Borgland}, {Borne}, {Bosch}, {Boutigny},
  {Brackett}, {Bradshaw}, {Brandt}, {Brown}, {Bullock}, {Burchat}, {Burke},
  {Cagnoli}, {Calabrese}, {Callahan}, {Callen}, {Carlin}, {Carlson},
  {Chandrasekharan}, {Charles-Emerson}, {Chesley}, {Cheu}, {Chiang}, {Chiang},
  {Chirino}, {Chow}, {Ciardi}, {Claver}, {Cohen-Tanugi}, {Cockrum}, {Coles},
  {Connolly}, {Cook}, {Cooray}, {Covey}, {Cribbs}, {Cui}, {Cutri}, {Daly},
  {Daniel}, {Daruich}, {Daubard}, {Daues}, {Dawson}, {Delgado}, {Dellapenna},
  {de Peyster}, {de Val-Borro}, {Digel}, {Doherty}, {Dubois},
  {Dubois-Felsmann}, {Durech}, {Economou}, {Eifler}, {Eracleous}, {Emmons},
  {Fausti Neto}, {Ferguson}, {Figueroa}, {Fisher-Levine}, {Focke}, {Foss},
  {Frank}, {Freemon}, {Gangler}, {Gawiser}, {Geary}, {Gee}, {Geha}, {Gessner},
  {Gibson}, {Gilmore}, {Glanzman}, {Glick}, {Goldina}, {Goldstein}, {Goodenow},
  {Graham}, {Gressler}, {Gris}, {Guy}, {Guyonnet}, {Haller}, {Harris},
  {Hascall}, {Haupt}, {Hernandez}, {Herrmann}, {Hileman}, {Hoblitt}, {Hodgson},
  {Hogan}, {Howard}, {Huang}, {Huffer}, {Ingraham}, {Innes}, {Jacoby}, {Jain},
  {Jammes}, {Jee}, {Jenness}, {Jernigan}, {Jevremovi{\'c}}, {Johns}, {Johnson},
  {Johnson}, {Jones}, {Juramy-Gilles}, {Juri{\'c}}, {Kalirai}, {Kallivayalil},
  {Kalmbach}, {Kantor}, {Karst}, {Kasliwal}, {Kelly}, {Kessler}, {Kinnison},
  {Kirkby}, {Knox}, {Kotov}, {Krabbendam}, {Krughoff}, {Kub{\'a}nek},
  {Kuczewski}, {Kulkarni}, {Ku}, {Kurita}, {Lage}, {Lambert}, {Lange},
  {Langton}, {Le Guillou}, {Levine}, {Liang}, {Lim}, {Lintott}, {Long},
  {Lopez}, {Lotz}, {Lupton}, {Lust}, {MacArthur}, {Mahabal}, {Mandelbaum},
  {Markiewicz}, {Marsh}, {Marshall}, {Marshall}, {May}, {McKercher}, {McQueen},
  {Meyers}, {Migliore}, {Miller}, {Mills}, {Miraval}, {Moeyens}, {Moolekamp},
  {Monet}, {Moniez}, {Monkewitz}, {Montgomery}, {Morrison}, {Mueller},
  {Muller}, {Mu{\~n}oz Arancibia}, {Neill}, {Newbry}, {Nief}, {Nomerotski},
  {Nordby}, {O'Connor}, {Oliver}, {Olivier}, {Olsen}, {O'Mullane}, {Ortiz},
  {Osier}, {Owen}, {Pain}, {Palecek}, {Parejko}, {Parsons}, {Pease},
  {Peterson}, {Peterson}, {Petravick}, {Libby Petrick}, {Petry},
  {Pierfederici}, {Pietrowicz}, {Pike}, {Pinto}, {Plante}, {Plate}, {Plutchak},
  {Price}, {Prouza}, {Radeka}, {Rajagopal}, {Rasmussen}, {Regnault}, {Reil},
  {Reiss}, {Reuter}, {Ridgway}, {Riot}, {Ritz}, {Robinson}, {Roby}, {Roodman},
  {Rosing}, {Roucelle}, {Rumore}, {Russo}, {Saha}, {Sassolas}, {Schalk},
  {Schellart}, {Schindler}, {Schmidt}, {Schneider}, {Schneider}, {Schoening},
  {Schumacher}, {Schwamb}, {Sebag}, {Selvy}, {Sembroski}, {Seppala}, {Serio},
  {Serrano}, {Shaw}, {Shipsey}, {Sick}, {Silvestri}, {Slater}, {Smith},
  {Smith}, {Sobhani}, {Soldahl}, {Storrie-Lombardi}, {Stover}, {Strauss},
  {Street}, {Stubbs}, {Sullivan}, {Sweeney}, {Swinbank}, {Szalay}, {Takacs},
  {Tether}, {Thaler}, {Thayer}, {Thomas}, {Thornton}, {Thukral}, {Tice},
  {Trilling}, {Turri}, {Van Berg}, {Vanden Berk}, {Vetter}, {Virieux},
  {Vucina}, {Wahl}, {Walkowicz}, {Walsh}, {Walter}, {Wang}, {Wang}, {Warner},
  {Wiecha}, {Willman}, {Winters}, {Wittman}, {Wolff}, {Wood-Vasey}, {Wu},
  {Xin}, {Yoachim}, \& {Zhan}}]{Ivezic2019}
{Ivezi{\'c}}, {\v{Z}}., {Kahn}, S.~M., {Tyson}, J.~A., {et~al.} 2019, \apj,
  873, 111

\bibitem[{{Jia} {et~al.}(2021){Jia}, {Ning}, {Sun}, {Yang}, \& {Cai}}]{Jia2021}
{Jia}, P., {Ning}, R., {Sun}, R., {Yang}, X., \& {Cai}, D. 2021, \mnras, 501,
  291

\bibitem[{Kinakh {et~al.}(2024)Kinakh, Belousov, Qu{\'e}tant, Drozdova,
  Holotyak, Schaerer, \& Voloshynovskiy}]{Kinakh2024}
Kinakh, V., Belousov, Y., Qu{\'e}tant, G., {et~al.} 2024, Sensors, 24

\bibitem[{{Kingma} \& {Welling}(2013)}]{Kingma2013}
{Kingma}, D.~P. \& {Welling}, M. 2013, arXiv e-prints, arXiv:1312.6114

\bibitem[{{Lanusse} {et~al.}(2021){Lanusse}, {Mandelbaum}, {Ravanbakhsh}, {Li},
  {Freeman}, \& {P{\'o}czos}}]{Lanusse2021}
{Lanusse}, F., {Mandelbaum}, R., {Ravanbakhsh}, S., {et~al.} 2021, \mnras, 504,
  5543

\bibitem[{{Lanusse} {et~al.}(2019){Lanusse}, {Melchior}, \&
  {Moolekamp}}]{Lanusse2019}
{Lanusse}, F., {Melchior}, P., \& {Moolekamp}, F. 2019, arXiv e-prints,
  arXiv:1912.03980

\bibitem[{{Laureijs} {et~al.}(2011){Laureijs}, {Amiaux}, {Arduini},
  {Augu{\`e}res}, {Brinchmann}, {Cole}, {Cropper}, {Dabin}, {Duvet}, {Ealet},
  {Garilli}, {Gondoin}, {Guzzo}, {Hoar}, {Hoekstra}, {Holmes}, {Kitching},
  {Maciaszek}, {Mellier}, {Pasian}, {Percival}, {Rhodes}, {Saavedra Criado},
  {Sauvage}, {Scaramella}, {Valenziano}, {Warren}, {Bender}, {Castander},
  {Cimatti}, {Le F{\`e}vre}, {Kurki-Suonio}, {Levi}, {Lilje}, {Meylan},
  {Nichol}, {Pedersen}, {Popa}, {Rebolo Lopez}, {Rix}, {Rottgering},
  {Zeilinger}, {Grupp}, {Hudelot}, {Massey}, {Meneghetti}, {Miller}, {Paltani},
  {Paulin-Henriksson}, {Pires}, {Saxton}, {Schrabback}, {Seidel}, {Walsh},
  {Aghanim}, {Amendola}, {Bartlett}, {Baccigalupi}, {Beaulieu}, {Benabed},
  {Cuby}, {Elbaz}, {Fosalba}, {Gavazzi}, {Helmi}, {Hook}, {Irwin}, {Kneib},
  {Kunz}, {Mannucci}, {Moscardini}, {Tao}, {Teyssier}, {Weller}, {Zamorani},
  {Zapatero Osorio}, {Boulade}, {Foumond}, {Di Giorgio}, {Guttridge}, {James},
  {Kemp}, {Martignac}, {Spencer}, {Walton}, {Bl{\"u}mchen}, {Bonoli},
  {Bortoletto}, {Cerna}, {Corcione}, {Fabron}, {Jahnke}, {Ligori}, {Madrid},
  {Martin}, {Morgante}, {Pamplona}, {Prieto}, {Riva}, {Toledo}, {Trifoglio},
  {Zerbi}, {Abdalla}, {Douspis}, {Grenet}, {Borgani}, {Bouwens}, {Courbin},
  {Delouis}, {Dubath}, {Fontana}, {Frailis}, {Grazian}, {Koppenh{\"o}fer},
  {Mansutti}, {Melchior}, {Mignoli}, {Mohr}, {Neissner}, {Noddle}, {Poncet},
  {Scodeggio}, {Serrano}, {Shane}, {Starck}, {Surace}, {Taylor},
  {Verdoes-Kleijn}, {Vuerli}, {Williams}, {Zacchei}, {Altieri}, {Escudero
  Sanz}, {Kohley}, {Oosterbroek}, {Astier}, {Bacon}, {Bardelli}, {Baugh},
  {Bellagamba}, {Benoist}, {Bianchi}, {Biviano}, {Branchini}, {Carbone},
  {Cardone}, {Clements}, {Colombi}, {Conselice}, {Cresci}, {Deacon}, {Dunlop},
  {Fedeli}, {Fontanot}, {Franzetti}, {Giocoli}, {Garcia-Bellido}, {Gow},
  {Heavens}, {Hewett}, {Heymans}, {Holland}, {Huang}, {Ilbert}, {Joachimi},
  {Jennins}, {Kerins}, {Kiessling}, {Kirk}, {Kotak}, {Krause}, {Lahav}, {van
  Leeuwen}, {Lesgourgues}, {Lombardi}, {Magliocchetti}, {Maguire}, {Majerotto},
  {Maoli}, {Marulli}, {Maurogordato}, {McCracken}, {McLure}, {Melchiorri},
  {Merson}, {Moresco}, {Nonino}, {Norberg}, {Peacock}, {Pello}, {Penny},
  {Pettorino}, {Di Porto}, {Pozzetti}, {Quercellini}, {Radovich}, {Rassat},
  {Roche}, {Ronayette}, {Rossetti}, {Sartoris}, {Schneider}, {Semboloni},
  {Serjeant}, {Simpson}, {Skordis}, {Smadja}, {Smartt}, {Spano}, {Spiro},
  {Sullivan}, {Tilquin}, {Trotta}, {Verde}, {Wang}, {Williger}, {Zhao},
  {Zoubian}, \& {Zucca}}]{Laureijs2011}
{Laureijs}, R., {Amiaux}, J., {Arduini}, S., {et~al.} 2011, arXiv e-prints,
  arXiv:1110.3193

\bibitem[{Ledig {et~al.}(2017)Ledig, Theis, Husz{\'a}r, Caballero, Cunningham,
  Acosta, Aitken, Tejani, Totz, Wang, {et~al.}}]{Ledig2017}
Ledig, C., Theis, L., Husz{\'a}r, F., {et~al.} 2017, in Proceedings of the IEEE
  conference on computer vision and pattern recognition, 4681--4690

\bibitem[{{Li} \& {Alexander}(2023)}]{Li2023}
{Li}, T. \& {Alexander}, E. 2023, \mnras, 522, L31

\bibitem[{{Li} {et~al.}(2024){Li}, {Do}, {Jones}, {Boscoe}, {Alfaro}, \&
  {Nguyen}}]{LiYunQi2024}
{Li}, Y.~Q., {Do}, T., {Jones}, E., {et~al.} 2024, arXiv e-prints,
  arXiv:2407.07229

\bibitem[{Liang {et~al.}(2021)Liang, Cao, Sun, Zhang, Van~Gool, \&
  Timofte}]{2021swinir}
Liang, J., Cao, J., Sun, G., {et~al.} 2021, in Proceedings of the IEEE/CVF
  international conference on computer vision, 1833--1844

\bibitem[{Lin {et~al.}(2021)Lin, Fouchez, \& Pasquet}]{Lin2021}
Lin, Q., Fouchez, D., \& Pasquet, J. 2021, in 2020 25th International
  Conference on Pattern Recognition (ICPR), 5634--5641

\bibitem[{{Lin} {et~al.}(2022){Lin}, {Fouchez}, {Pasquet}, {Treyer}, {Ait
  Ouahmed}, {Arnouts}, \& {Ilbert}}]{Lin2022}
{Lin}, Q., {Fouchez}, D., {Pasquet}, J., {et~al.} 2022, \aap, 662, A36

\bibitem[{{Lin} {et~al.}(2024){Lin}, {Ruan}, {Fouchez}, {Chen}, {Li},
  {Montero-Camacho}, {Napolitano}, {Ting}, \& {Zhang}}]{Lin2024}
{Lin}, Q., {Ruan}, H., {Fouchez}, D., {et~al.} 2024, \aap, 691, A331

\bibitem[{{Lintott} {et~al.}(2011){Lintott}, {Schawinski}, {Bamford}, {Slosar},
  {Land}, {Thomas}, {Edmondson}, {Masters}, {Nichol}, {Raddick}, {Szalay},
  {Andreescu}, {Murray}, \& {Vandenberg}}]{Lintott2011}
{Lintott}, C., {Schawinski}, K., {Bamford}, S., {et~al.} 2011, \mnras, 410, 166

\bibitem[{Liu {et~al.}(2021)Liu, Lin, Cao, Hu, Wei, Zhang, Lin, \&
  Guo}]{SwinTransformer}
Liu, Z., Lin, Y., Cao, Y., {et~al.} 2021, in 2021 IEEE/CVF International
  Conference on Computer Vision (ICCV), 9992--10002

\bibitem[{{Lizarraga} {et~al.}(2024){Lizarraga}, {Hanchen Jiang}, {Nowack},
  {Li}, {Nian Wu}, {Boscoe}, \& {Do}}]{Lizarraga2024}
{Lizarraga}, A., {Hanchen Jiang}, E., {Nowack}, J., {et~al.} 2024, arXiv
  e-prints, arXiv:2411.18440

\bibitem[{{Luo} {et~al.}(2024){Luo}, {Tang}, {Chen}, {Fu}, {Du}, {Zhang},
  {Gong}, {Shu}, {Lu}, {Li}, {Meng}, {Zhou}, \& {Fan}}]{Luo2024}
{Luo}, Z., {Tang}, Z., {Chen}, Z., {et~al.} 2024, \mnras, 531, 3539

\bibitem[{{Luo} {et~al.}(2025){Luo}, {Zhang}, {Chen}, {Chen}, {Fu}, {Xiao},
  {Du}, \& {Shu}}]{Luo2025}
{Luo}, Z., {Zhang}, S., {Chen}, J., {et~al.} 2025, \apjs, 277, 22

\bibitem[{{Lupton} {et~al.}(2004){Lupton}, {Blanton}, {Fekete}, {Hogg},
  {O'Mullane}, {Szalay}, \& {Wherry}}]{Lupton2004}
{Lupton}, R., {Blanton}, M.~R., {Fekete}, G., {et~al.} 2004, \pasp, 116, 133

\bibitem[{{Miao} {et~al.}(2024){Miao}, {Tu}, {Jiang}, {Li}, \&
  {Qiu}}]{Miao2024}
{Miao}, J., {Tu}, L., {Jiang}, B., {Li}, X., \& {Qiu}, B. 2024, \apjs, 274, 7

\bibitem[{Miao {et~al.}(2025)Miao, Tu, Liu, \& Zhao}]{Miao2025}
Miao, J., Tu, L., Liu, H., \& Zhao, J. 2025, \apjs, 278, 35

\bibitem[{{Papamakarios} {et~al.}(2017){Papamakarios}, {Pavlakou}, \&
  {Murray}}]{Papamakarios2017}
{Papamakarios}, G., {Pavlakou}, T., \& {Murray}, I. 2017, arXiv e-prints,
  arXiv:1705.07057

\bibitem[{{Park} {et~al.}(2024){Park}, {Jo}, {Kang}, {Kim}, \&
  {Jee}}]{Park2024}
{Park}, H., {Jo}, Y., {Kang}, S., {Kim}, T., \& {Jee}, M.~J. 2024, \apj, 972,
  45

\bibitem[{{Pasquet} {et~al.}(2019){Pasquet}, {Bertin}, {Treyer}, {Arnouts}, \&
  {Fouchez}}]{Pasquet2019}
{Pasquet}, J., {Bertin}, E., {Treyer}, M., {Arnouts}, S., \& {Fouchez}, D.
  2019, \aap, 621, A26

\bibitem[{{Ravanbakhsh} {et~al.}(2016){Ravanbakhsh}, {Lanusse}, {Mandelbaum},
  {Schneider}, \& {Poczos}}]{Ravanbakhsh2016}
{Ravanbakhsh}, S., {Lanusse}, F., {Mandelbaum}, R., {Schneider}, J., \&
  {Poczos}, B. 2016, arXiv e-prints, arXiv:1609.05796

\bibitem[{{Reiman} \& {G{\"o}hre}(2019)}]{Reiman2019}
{Reiman}, D.~M. \& {G{\"o}hre}, B.~E. 2019, \mnras, 485, 2617

\bibitem[{Rezende \& Mohamed(2015)}]{Rezende2015}
Rezende, D.~J. \& Mohamed, S. 2015, in Proceedings of the 32nd International
  Conference on International Conference on Machine Learning - Volume 37,
  ICML'15 (JMLR.org), 1530–1538

\bibitem[{Ronneberger {et~al.}(2015)Ronneberger, Fischer, \&
  Brox}]{Ronneberger2015}
Ronneberger, O., Fischer, P., \& Brox, T. 2015, in Medical Image Computing and
  Computer-Assisted Intervention -- MICCAI 2015, ed. N.~Navab, J.~Hornegger,
  W.~M. Wells, \& A.~F. Frangi (Cham: Springer International Publishing),
  234--241

\bibitem[{Sabour {et~al.}(2017)Sabour, Frosst, \& Hinton}]{Sabour2017}
Sabour, S., Frosst, N., \& Hinton, G.~E. 2017, in Proceedings of the 31st
  International Conference on Neural Information Processing Systems, NIPS'17
  (Red Hook, NY, USA: Curran Associates Inc.), 3859–3869

\bibitem[{Saharia {et~al.}(2022)Saharia, Chan, Chang, Lee, Ho, Salimans, Fleet,
  \& Norouzi}]{Saharia2022}
Saharia, C., Chan, W., Chang, H., {et~al.} 2022, in ACM SIGGRAPH 2022
  Conference Proceedings, SIGGRAPH '22 (New York, NY, USA: Association for
  Computing Machinery)

\bibitem[{Salimans {et~al.}(2016)Salimans, Goodfellow, Zaremba, Cheung,
  Radford, \& Chen}]{Salimans2016}
Salimans, T., Goodfellow, I., Zaremba, W., {et~al.} 2016, in Proceedings of the
  30th International Conference on Neural Information Processing Systems,
  NIPS'16 (Red Hook, NY, USA: Curran Associates Inc.), 2234–2242

\bibitem[{{Salimans} {et~al.}(2017){Salimans}, {Karpathy}, {Chen}, \&
  {Kingma}}]{Salimans2017}
{Salimans}, T., {Karpathy}, A., {Chen}, X., \& {Kingma}, D.~P. 2017, arXiv
  e-prints, arXiv:1701.05517

\bibitem[{{Schawinski} {et~al.}(2017){Schawinski}, {Zhang}, {Zhang}, {Fowler},
  \& {Santhanam}}]{Schawinski2017}
{Schawinski}, K., {Zhang}, C., {Zhang}, H., {Fowler}, L., \& {Santhanam}, G.~K.
  2017, \mnras, 467, L110

\bibitem[{{Schlegel} {et~al.}(1998){Schlegel}, {Finkbeiner}, \&
  {Davis}}]{Schlegel1998}
{Schlegel}, D.~J., {Finkbeiner}, D.~P., \& {Davis}, M. 1998, \apj, 500, 525

\bibitem[{Shan {et~al.}(2025)Shan, Liu, Qiu, Luo, ji~Ren, Pan, \&
  Chen}]{Shan2025}
Shan, Q.-Q., Liu, C.-X., Qiu, B., {et~al.} 2025, Engineering Applications of
  Artificial Intelligence, 142, 109836

\bibitem[{{Shibuya} {et~al.}(2025){Shibuya}, {Ito}, {Asai}, {Kirihara},
  {Fujimoto}, {Toba}, {Miura}, {Umayahara}, {Iwadate}, {Ali}, \&
  {Kodama}}]{Shibuya2025}
{Shibuya}, T., {Ito}, Y., {Asai}, K., {et~al.} 2025, \pasj, 77, 21

\bibitem[{{Smith} {et~al.}(2022){Smith}, {Geach}, {Jackson}, {Arora}, {Stone},
  \& {Courteau}}]{Smith2022}
{Smith}, M.~J., {Geach}, J.~E., {Jackson}, R.~A., {et~al.} 2022, \mnras, 511,
  1808

\bibitem[{Song {et~al.}(2021)Song, Meng, \& Ermon}]{Song2021}
Song, J., Meng, C., \& Ermon, S. 2021, in International Conference on Learning
  Representations

\bibitem[{{Spergel} {et~al.}(2015){Spergel}, {Gehrels}, {Baltay}, {Bennett},
  {Breckinridge}, {Donahue}, {Dressler}, {Gaudi}, {Greene}, {Guyon}, {Hirata},
  {Kalirai}, {Kasdin}, {Macintosh}, {Moos}, {Perlmutter}, {Postman},
  {Rauscher}, {Rhodes}, {Wang}, {Weinberg}, {Benford}, {Hudson}, {Jeong},
  {Mellier}, {Traub}, {Yamada}, {Capak}, {Colbert}, {Masters}, {Penny},
  {Savransky}, {Stern}, {Zimmerman}, {Barry}, {Bartusek}, {Carpenter}, {Cheng},
  {Content}, {Dekens}, {Demers}, {Grady}, {Jackson}, {Kuan}, {Kruk}, {Melton},
  {Nemati}, {Parvin}, {Poberezhskiy}, {Peddie}, {Ruffa}, {Wallace}, {Whipple},
  {Wollack}, \& {Zhao}}]{Spergel2015}
{Spergel}, D., {Gehrels}, N., {Baltay}, C., {et~al.} 2015, arXiv e-prints,
  arXiv:1503.03757

\bibitem[{{Spindler} {et~al.}(2021){Spindler}, {Geach}, \&
  {Smith}}]{Spindler2021}
{Spindler}, A., {Geach}, J.~E., \& {Smith}, M.~J. 2021, \mnras, 502, 985

\bibitem[{{Storey-Fisher} {et~al.}(2021){Storey-Fisher}, {Huertas-Company},
  {Ramachandra}, {Lanusse}, {Leauthaud}, {Luo}, {Huang}, \&
  {Prochaska}}]{StoreyFisher2021}
{Storey-Fisher}, K., {Huertas-Company}, M., {Ramachandra}, N., {et~al.} 2021,
  \mnras, 508, 2946

\bibitem[{{Ting}(2025)}]{Ting2025}
{Ting}, Y.-S. 2025, The Open Journal of Astrophysics, 8, 95

\bibitem[{{Uria} {et~al.}(2016){Uria}, {C{\^o}t{\'e}}, {Gregor}, {Murray}, \&
  {Larochelle}}]{Uria2016}
{Uria}, B., {C{\^o}t{\'e}}, M.-A., {Gregor}, K., {Murray}, I., \& {Larochelle},
  H. 2016, arXiv e-prints, arXiv:1605.02226

\bibitem[{van~den Oord {et~al.}(2016)van~den Oord, Kalchbrenner, \&
  Kavukcuoglu}]{vandenOord2016}
van~den Oord, A., Kalchbrenner, N., \& Kavukcuoglu, K. 2016, in Proceedings of
  Machine Learning Research, Vol.~48, Proceedings of The 33rd International
  Conference on Machine Learning, ed. M.~F. Balcan \& K.~Q. Weinberger (New
  York, New York, USA: PMLR), 1747--1756

\bibitem[{Vaswani {et~al.}(2017)Vaswani, Shazeer, Parmar, Uszkoreit, Jones,
  Gomez, Kaiser, \& Polosukhin}]{Vaswani2017}
Vaswani, A., Shazeer, N., Parmar, N., {et~al.} 2017, in Proceedings of the 31st
  International Conference on Neural Information Processing Systems, NIPS'17
  (Red Hook, NY, USA: Curran Associates Inc.), 6000–6010

\bibitem[{{Wang} {et~al.}(2023){Wang}, {Sreejith}, {Lin}, {Ramachandra},
  {Solsar}, \& {Yoo}}]{Wang2023}
{Wang}, H., {Sreejith}, S., {Lin}, Y., {et~al.} 2023, The Open Journal of
  Astrophysics, 6, 30

\bibitem[{{Wang} {et~al.}(2022){Wang}, {Sreejith}, {Slosar}, {Lin}, \&
  {Yoo}}]{Wang2022}
{Wang}, H., {Sreejith}, S., {Slosar}, A., {Lin}, Y., \& {Yoo}, S. 2022, \prd,
  106, 063023

\bibitem[{Wang {et~al.}(2022)Wang, Xie, Yu, Chan, Loy, \& Dong}]{2022basicsr}
Wang, X., Xie, L., Yu, K., {et~al.} 2022, {BasicSR}: Open Source Image and
  Video Restoration Toolbox, \url{https://github.com/XPixelGroup/BasicSR}

\bibitem[{Wang {et~al.}(2004)Wang, Bovik, Sheikh, \& Simoncelli}]{Wang2004}
Wang, Z., Bovik, A., Sheikh, H., \& Simoncelli, E. 2004, IEEE Transactions on
  Image Processing, 13, 600

\bibitem[{{York} {et~al.}(2000){York}, {Adelman}, {Anderson}, {Anderson},
  {Annis}, {Bahcall}, {Bakken}, {Barkhouser}, {Bastian}, {Berman}, {Boroski},
  {Bracker}, {Briegel}, {Briggs}, {Brinkmann}, {Brunner}, {Burles}, {Carey},
  {Carr}, {Castander}, {Chen}, {Colestock}, {Connolly}, {Crocker}, {Csabai},
  {Czarapata}, {Davis}, {Doi}, {Dombeck}, {Eisenstein}, {Ellman}, {Elms},
  {Evans}, {Fan}, {Federwitz}, {Fiscelli}, {Friedman}, {Frieman}, {Fukugita},
  {Gillespie}, {Gunn}, {Gurbani}, {de Haas}, {Haldeman}, {Harris}, {Hayes},
  {Heckman}, {Hennessy}, {Hindsley}, {Holm}, {Holmgren}, {Huang}, {Hull},
  {Husby}, {Ichikawa}, {Ichikawa}, {Ivezi{\'c}}, {Kent}, {Kim}, {Kinney},
  {Klaene}, {Kleinman}, {Kleinman}, {Knapp}, {Korienek}, {Kron}, {Kunszt},
  {Lamb}, {Lee}, {Leger}, {Limmongkol}, {Lindenmeyer}, {Long}, {Loomis},
  {Loveday}, {Lucinio}, {Lupton}, {MacKinnon}, {Mannery}, {Mantsch}, {Margon},
  {McGehee}, {McKay}, {Meiksin}, {Merelli}, {Monet}, {Munn}, {Narayanan},
  {Nash}, {Neilsen}, {Neswold}, {Newberg}, {Nichol}, {Nicinski}, {Nonino},
  {Okada}, {Okamura}, {Ostriker}, {Owen}, {Pauls}, {Peoples}, {Peterson},
  {Petravick}, {Pier}, {Pope}, {Pordes}, {Prosapio}, {Rechenmacher}, {Quinn},
  {Richards}, {Richmond}, {Rivetta}, {Rockosi}, {Ruthmansdorfer}, {Sandford},
  {Schlegel}, {Schneider}, {Sekiguchi}, {Sergey}, {Shimasaku}, {Siegmund},
  {Smee}, {Smith}, {Snedden}, {Stone}, {Stoughton}, {Strauss}, {Stubbs},
  {SubbaRao}, {Szalay}, {Szapudi}, {Szokoly}, {Thakar}, {Tremonti}, {Tucker},
  {Uomoto}, {Vanden Berk}, {Vogeley}, {Waddell}, {Wang}, {Watanabe},
  {Weinberg}, {Yanny}, {Yasuda}, \& {SDSS Collaboration}}]{York2000}
{York}, D.~G., {Adelman}, J., {Anderson}, Jr., J.~E., {et~al.} 2000, \aj, 120,
  1579

\bibitem[{{Zanisi} {et~al.}(2021){Zanisi}, {Huertas-Company}, {Lanusse},
  {Bottrell}, {Pillepich}, {Nelson}, {Rodriguez-Gomez}, {Shankar}, {Hernquist},
  {Dekel}, {Margalef-Bentabol}, {Vogelsberger}, \& {Primack}}]{Zanisi2021}
{Zanisi}, L., {Huertas-Company}, M., {Lanusse}, F., {et~al.} 2021, \mnras, 501,
  4359

\bibitem[{{Zhan}(2018)}]{Zhan2018}
{Zhan}, H. 2018, in 42nd COSPAR Scientific Assembly, Vol.~42, E1.16--4--18

\bibitem[{{Zhang} {et~al.}(2024){Zhang}, {Liu}, {Yi}, {Yuan}, {Yang}, {Bu},
  {Kong}, {Jia}, {Bi}, {Zhang}, \& {Li}}]{Zhang2024}
{Zhang}, R., {Liu}, M., {Yi}, Z., {et~al.} 2024, \pasa, 41, e035

\bibitem[{{Zhou} {et~al.}(2022){Zhou}, {Gu}, {Fang}, \& {Lin}}]{Zhou2022}
{Zhou}, C., {Gu}, Y., {Fang}, G., \& {Lin}, Z. 2022, \aj, 163, 86

\end{thebibliography}

\onecolumn

\appendix

\section{Network architectures} \label{sec:network}

\begin{figure*}[h]
\centering
\begin{subfigure}{0.8\textwidth}
\centering
\includegraphics[width=0.75\textwidth]{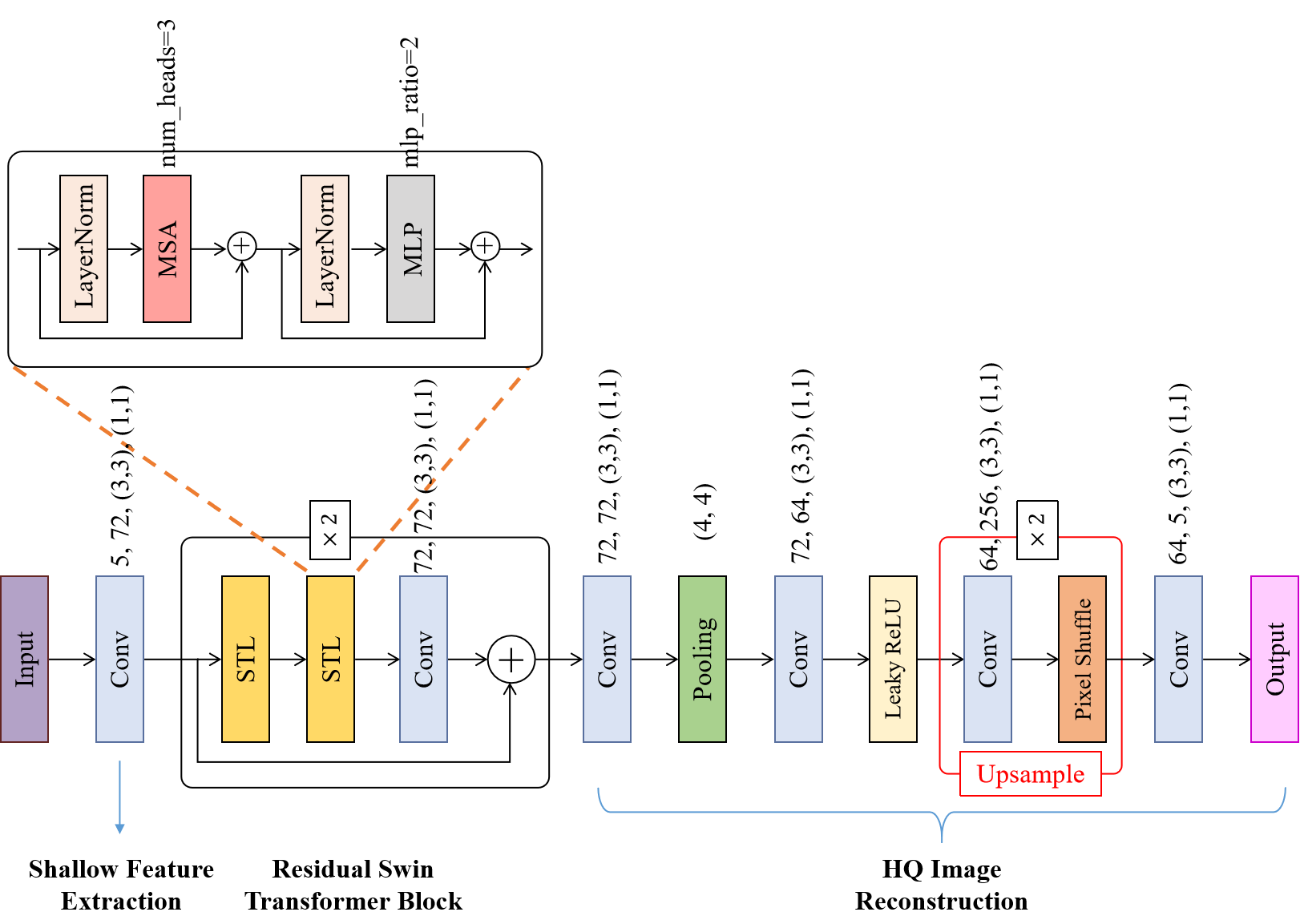}
\caption{}
\label{fig:swinir_s2s}
\end{subfigure}
\hfill
\begin{subfigure}{0.8\textwidth}
\centering
\includegraphics[width=0.75\textwidth]{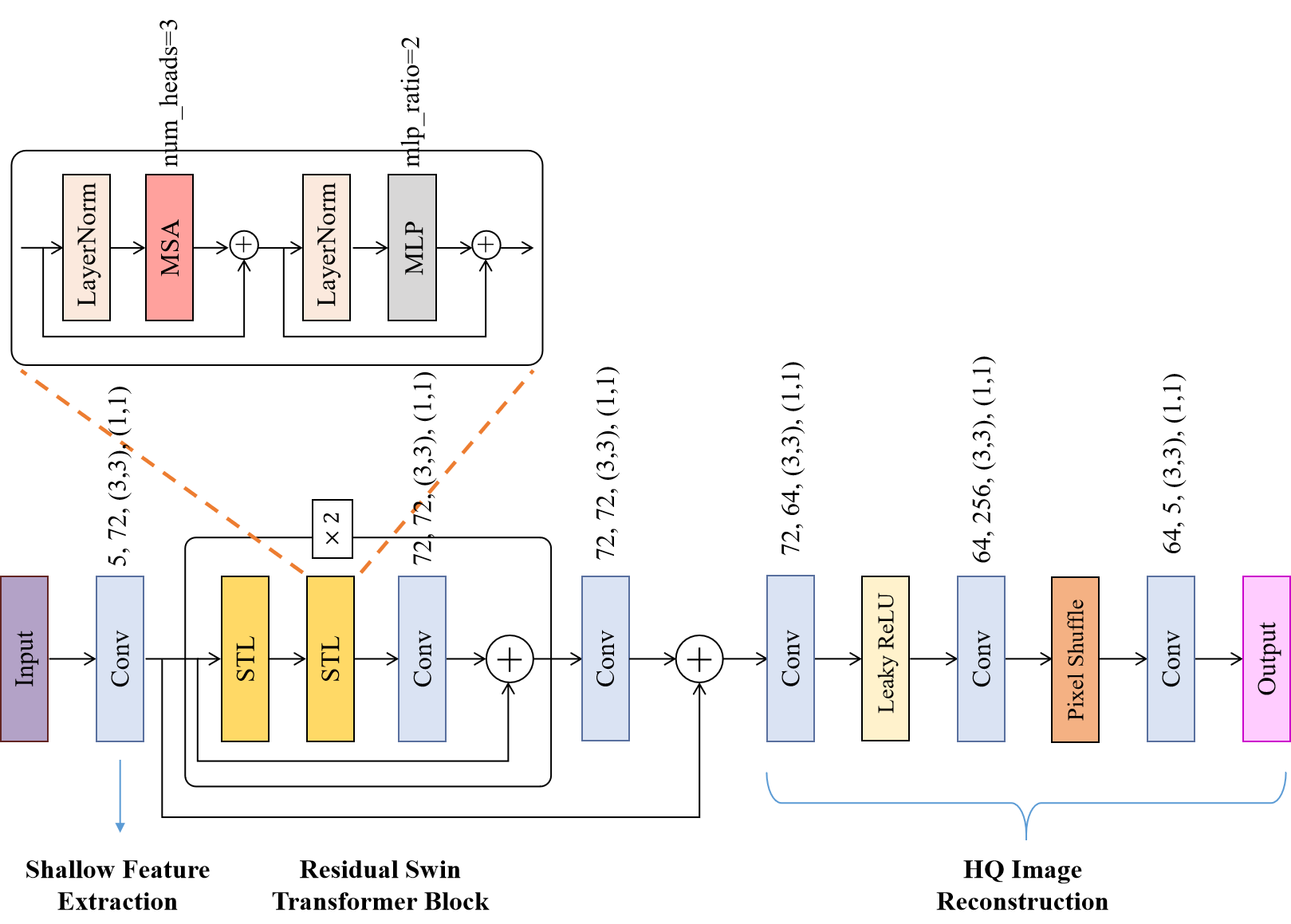}
\caption{}
\label{fig:swinir_s2c}
\end{subfigure}
\caption{Architectures of the Swin Transformer for the S2S and S2C translations, illustrated in subfigures (a) and (b), respectively. The three components of the Swin Transformer are shown, i.e. a shallow feature extraction module, residual Swin Transformer blocks (RSTBs), and a high-quality (HQ) image reconstruction module. The shallow feature extraction module is a convolutional layer. Two RSTBs are implemented, each composed of two Swin Transformer layers (STLs) and a convolutional layer. Each STL contains a multi-head self-attention (MSA) unit with three attention heads, two layer normalization (LayerNorm) operations, a multi-layer perceptron (MLP) with an MLP ratio of 2, and two skip connections. The HQ image reconstruction module mainly consists of convolutional layers. For the S2S translation, a $4\times4$ average pooling layer for downsampling and two pixel shuffle layers for upsampling are implemented. For the S2C translation, a pixel shuffle layer is implemented, and there is a skip connection that directly links the shallow feature extraction module to the HQ image reconstruction module. The blocks that are repeated multiple times are marked with ``$\times N$''. The leaky rectified linear unit (Leaky ReLU) activation has a leaky ratio of 0.01. The numbers next to each convolutional layer refer to the number of input channels, the number of output channels, the kernel size, and the stride. For instance, ``5, 72, (3,3), (1,1)'' means that a convolutional layer has five input channels and 72 output channels, in which $3\times3$ kernels and $1\times1$ strides are implemented. The ``same'' padding is applied in all the convolutional layers.}
\label{fig:swinir}
\end{figure*}

\begin{figure*}
\centering
\begin{subfigure}[b]{1\textwidth}
\centering
\includegraphics[width=0.8\textwidth]{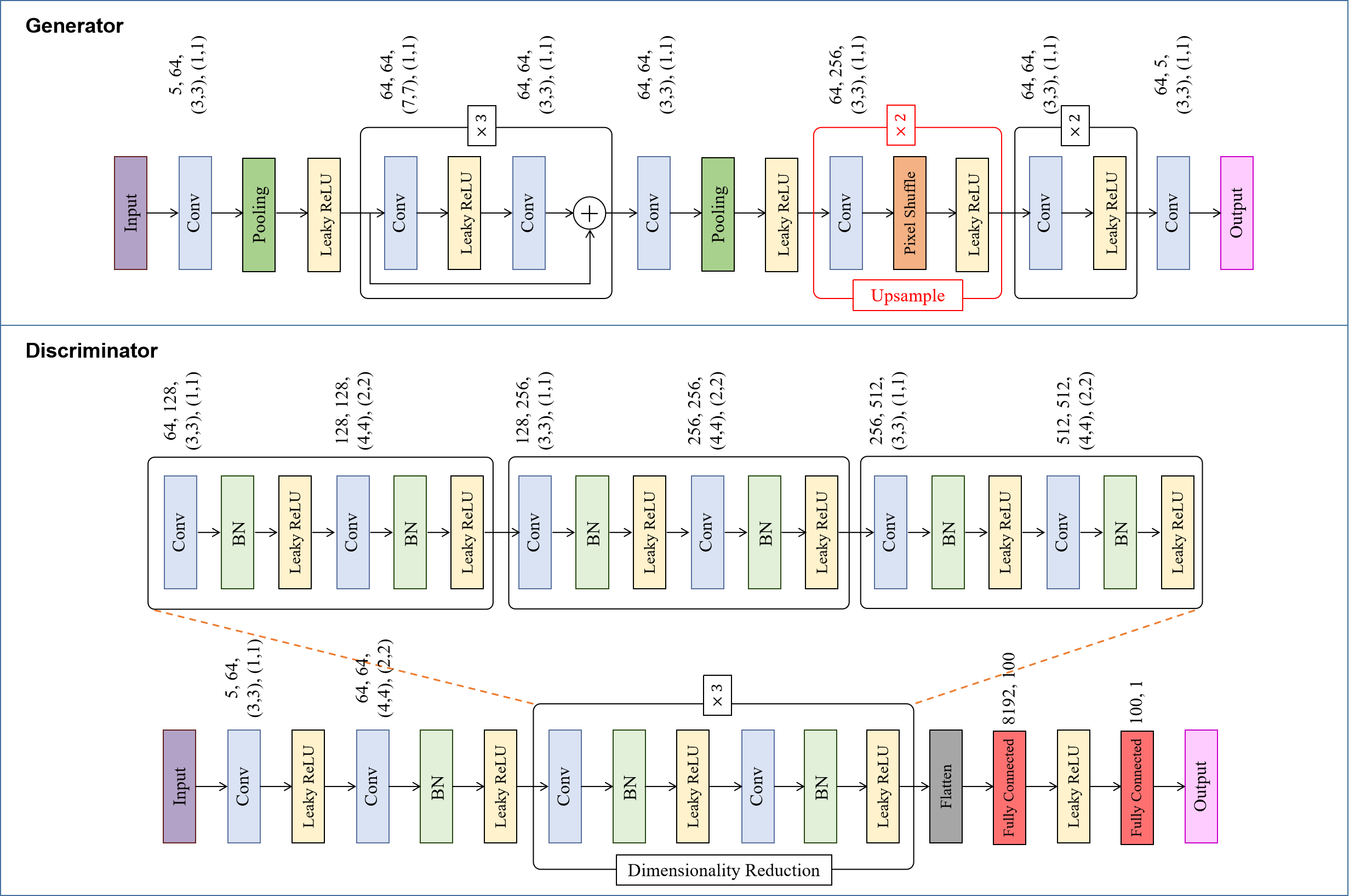}
\caption{}
\label{fig:srgan_s2s}
\end{subfigure}
\hfill
\begin{subfigure}{1\textwidth}
\centering
\includegraphics[width=0.8\textwidth]{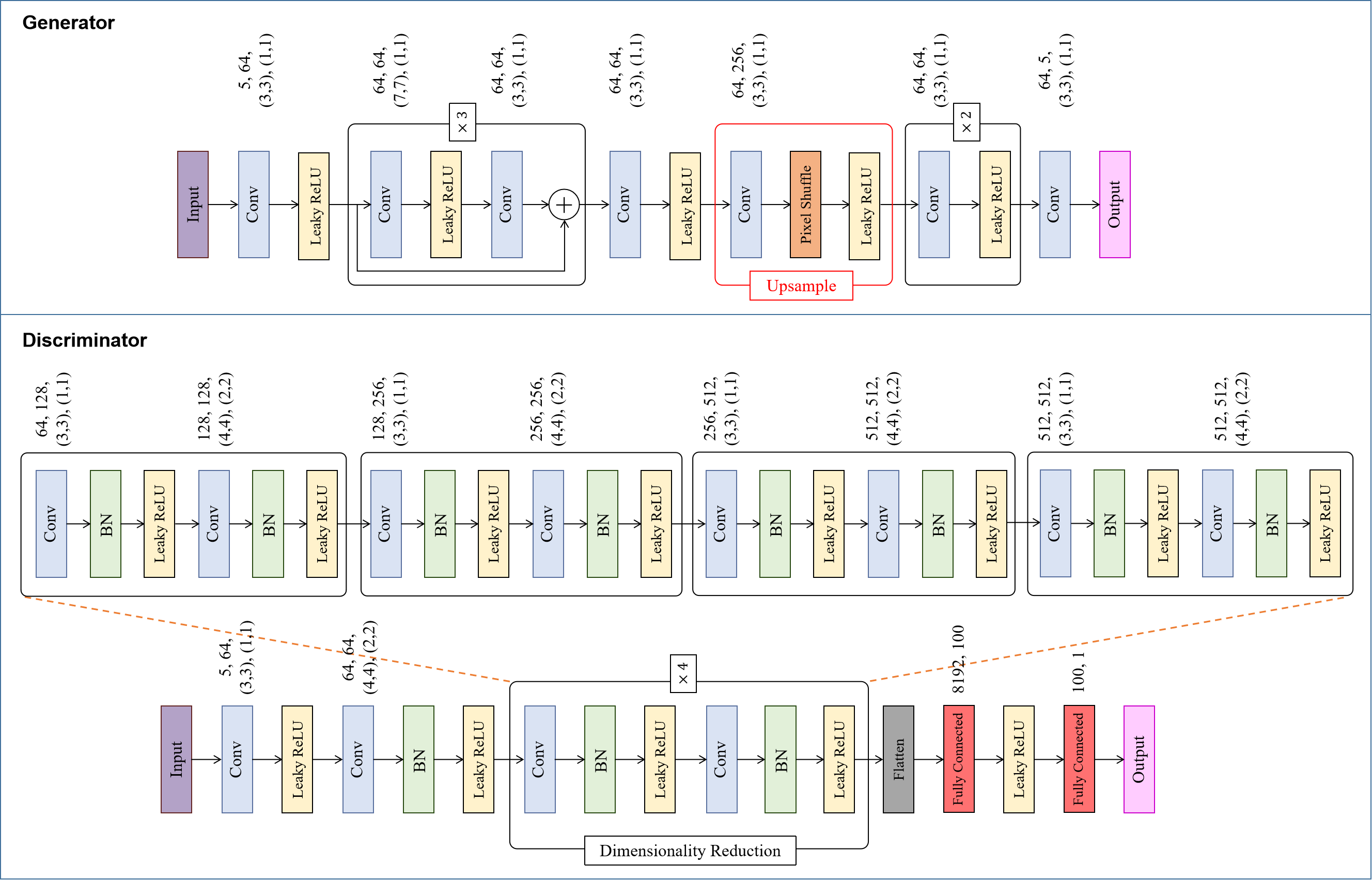}
\caption{}
\label{fig:srgan_s2c}
\end{subfigure}
\caption{Architectures of the SRGAN for the S2S and S2C translations, illustrated in subfigures (a) and (b), respectively. The generator and discriminator architectures both mainly consist of convolutional layers and batch normalization (BN) operations. Two $2\times2$ average pooling layers for downsampling and two pixel shuffle layers for upsampling are implemented in the generator for the S2S translation, while a pixel shuffle layer is implemented for the S2C translation. There are also skip connections in the two generators. The blocks that are repeated multiple times are marked with ``$\times N$''. The Leaky ReLU activation applied in the generators and discriminators has leaky ratios of 0.01 and 0.2, respectively. Same as Fig.~\ref{fig:swinir}, the numbers next to each convolutional layer refer to the number of input channels, the number of output channels, the kernel size, and the stride. The ``same'' padding is applied in all the convolutional layers. The stride-2 convolutional layers are implemented in the discriminators for dimensionality reduction, followed by flattening and fully connected layers. The numbers next to each fully connected layer refer to the number of input dimensions and the number of output dimensions.}
\label{fig:srgan}
\end{figure*}

\begin{figure*}
\centering
\includegraphics[width=0.9\textwidth]{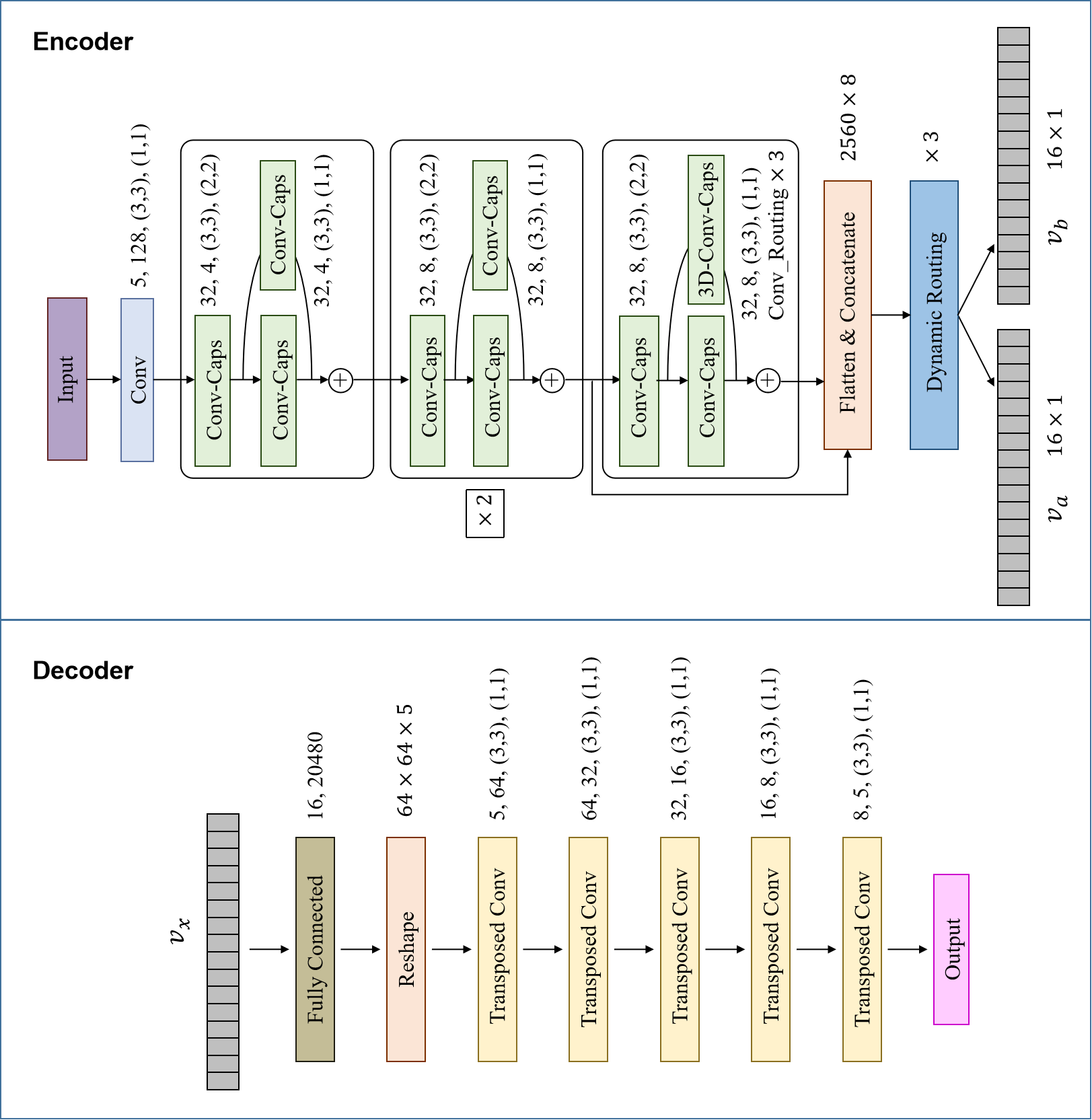}
\caption{Architecture of the capsule network for the S2S translation. The encoder consists of a convolutional layer, four capsule modules, flattening, concatenation, and routing operations. The convolutional layer has five input channels, 128 output channels, $3\times3$ kernels, $1\times1$ strides, and is followed by the ReLU activation and batch normalization. Each of the first three capsule modules contains three convolutional capsule layers (Conv-Caps) with skip connections. The numbers next to each convolutional capsule layer refer to the number of capsules, the number of dimensions for each capsule, the convolutional kernel size, and the stride. The last capsule module has a three-dimensional convolutional capsule layer (3D-Conv-Caps), in which the convolution-based routing is applied three times. The spatial dimensions are reduced by each capsule module due to the use of $2\times2$ strides. The outputs of the third and fourth capsule modules, which have $8\times8\times32\times8$ and $4\times4\times32\times8$ dimensions, respectively, are concatenated and flattened into a tensor of shape $2560\times8$. It is fed into the final layer that applies the dynamic routing three times and outputs two vectors of length 16 (i.e. $v_a$ and $v_b$). The decoder consists of a fully connected layer, a reshaping operation, and five transposed convolutional layers. It takes one of the vectors (i.e. $v_x$) as input. The fully connected layer has 16 input dimensions and 20480 output dimensions, whose output is then reshaped into $64\times64\times5$ dimensions. The numbers next to each transposed convolutional layer refer to the number of input channels, the number of output channels, the kernel size, and the stride. The fully connected layer and each of the first four transposed convolutional layers are followed by the Parametric ReLU (PReLU) activation. The ``same'' padding is applied in all the (transposed) convolutional layers. In the capsule network implementation in which the morphological classifications are not used, the encoder only outputs one vector and it is directly input to the decoder.}
\label{fig:capsule}
\end{figure*}

\begin{figure*}
\centering
\includegraphics[width=0.95\textwidth]{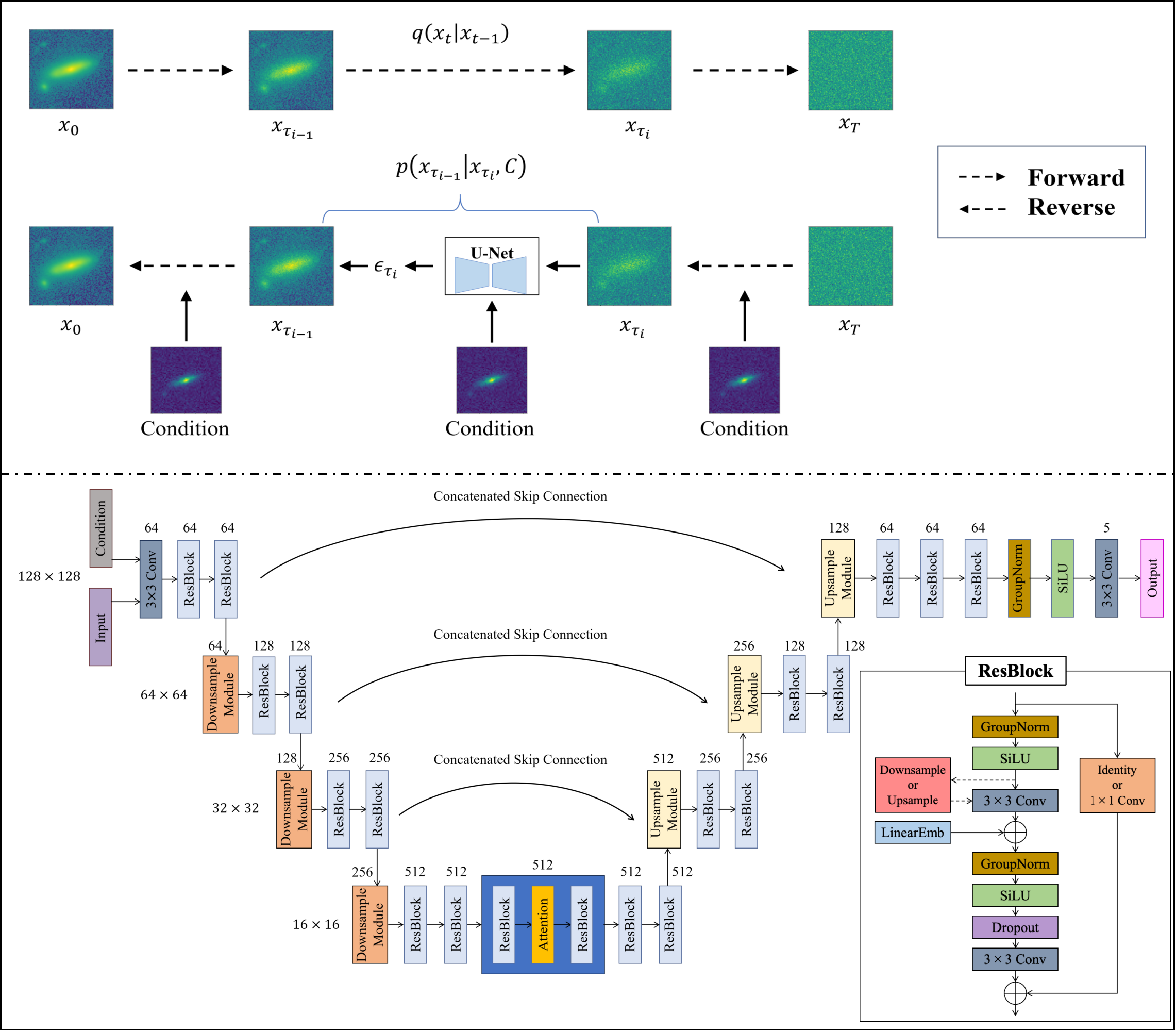}
\caption{Illustration of the diffusion model for the S2C translation. \textit{Upper panel:} Schematic illustration of the forward and reverse (generative) processes. The forward process corrupts the original target CFHTLS images from $x_0$ to $x_T$ with Gaussian noise, where $T$ is the total number of diffusion steps ($T=1000$ in this work). Each step is represented by $q(x_t | x_{t-1})$. The generative process takes a sub-sequence from the $T$ steps, in which each step is represented by $p(x_{\tau_{i-1}} | x_{\tau_i}, C)$ where $C$ represents the conditions (i.e. the SDSS images upsampled to $128\times128$ resolution via bicubic interpolation). It is realized via a U-Net that is fed with $x_{\tau_i}$ and $C$. The U-Net outputs a noise image, $\epsilon_{\tau_i}$, and $\epsilon_{\tau_i}$ is used to construct $x_{\tau_{i-1}}$. Through this iterative process, the noise-like images $x_T$ can be gradually denoised and approximate the target images $x_0$ given the conditions $C$. \textit{Lower panel:} Architecture of the U-Net. It mainly consists of residual blocks (ResBlocks), downsample modules, upsample modules, and a self-attention block. The first and the last layers of the U-Net are both convolutional layers with $3\times3$ kernels, $1\times1$ strides, and the ``same'' padding. A residual block (shown in the bottom-right corner) consists of group normalization (GroupNorm) operations, the sigmoid linear Unit (SiLU) activation, two $3\times3$ convolutional layers, a linear embedding, a dropout layer (with a dropout ratio of 0.2), and a skip connection. The linear embedding is applied for encoding the noise level that is used to corrupt the images. The skip connection is realized by a $1\times1$ convolution if the input and the output of the residual block do not have the same number of channels. The spatial dimensions of the input and the output are the same if neither downsampling nor upsampling is applied in the residual block. On the other hand, a downsample module has the same structure as a residual block but a $2\times2$ average pooling layer is implemented before the first $3\times3$ convolution. An upsample module contains two residual blocks, the second of which applies nearest neighbor interpolation before the first $3\times3$ convolution. Viewing the whole U-Net, the inputs of the first 12 residual blocks or downsample modules are also fed into the last 12 residual blocks or upsample modules in one-to-one correspondence in reverse order (via concatenation), as indicated by the curved arrows. The spatial dimensions of the outputs of all the blocks in each row remain the same. They are indicated by the numbers on the left of the rows. The number next to each block stands for the number of its output channels.}
\label{fig:diffu_s2c}
\end{figure*}

\FloatBarrier

\section{More results on image translation} \label{sec:res_more}

\begin{figure*}[h]
\begin{center}
\centerline{\includegraphics[width=0.65\linewidth]{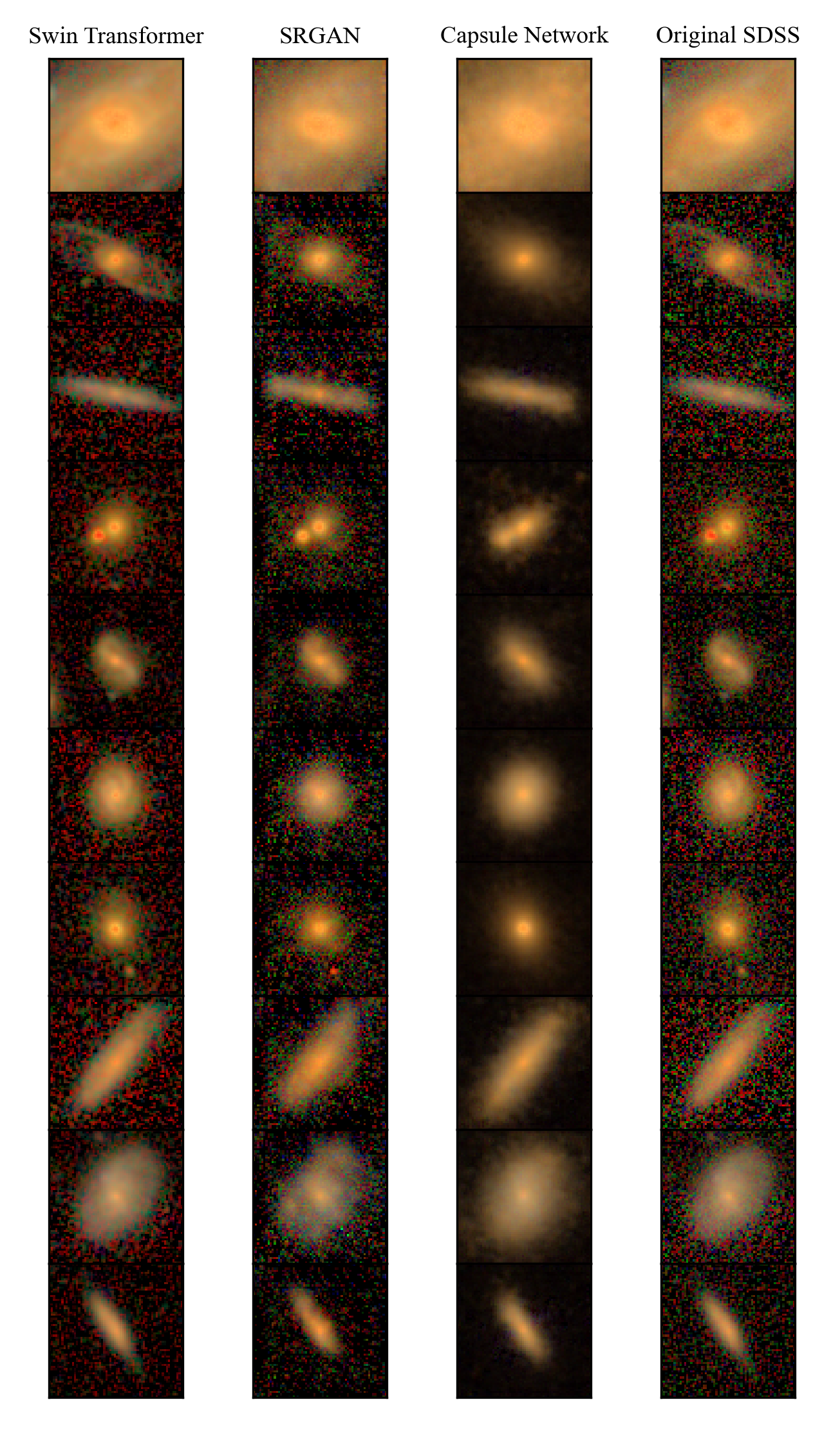}}
\caption{Exemplar RGB images generated by the S2S translation models (i.e. the Swin Transformer, the SRGAN, and the capsule network) and the corresponding original SDSS validation images. The images were first restored to the initial flux scale using the reverse of Eq.~\ref{eq:norm} and then converted to RGB format using the method from \citet{Lupton2004}.}
\label{fig:rgb_s2s}
\end{center}
\end{figure*}

\begin{figure*}
\begin{center}
\centerline{\includegraphics[width=0.85\linewidth]{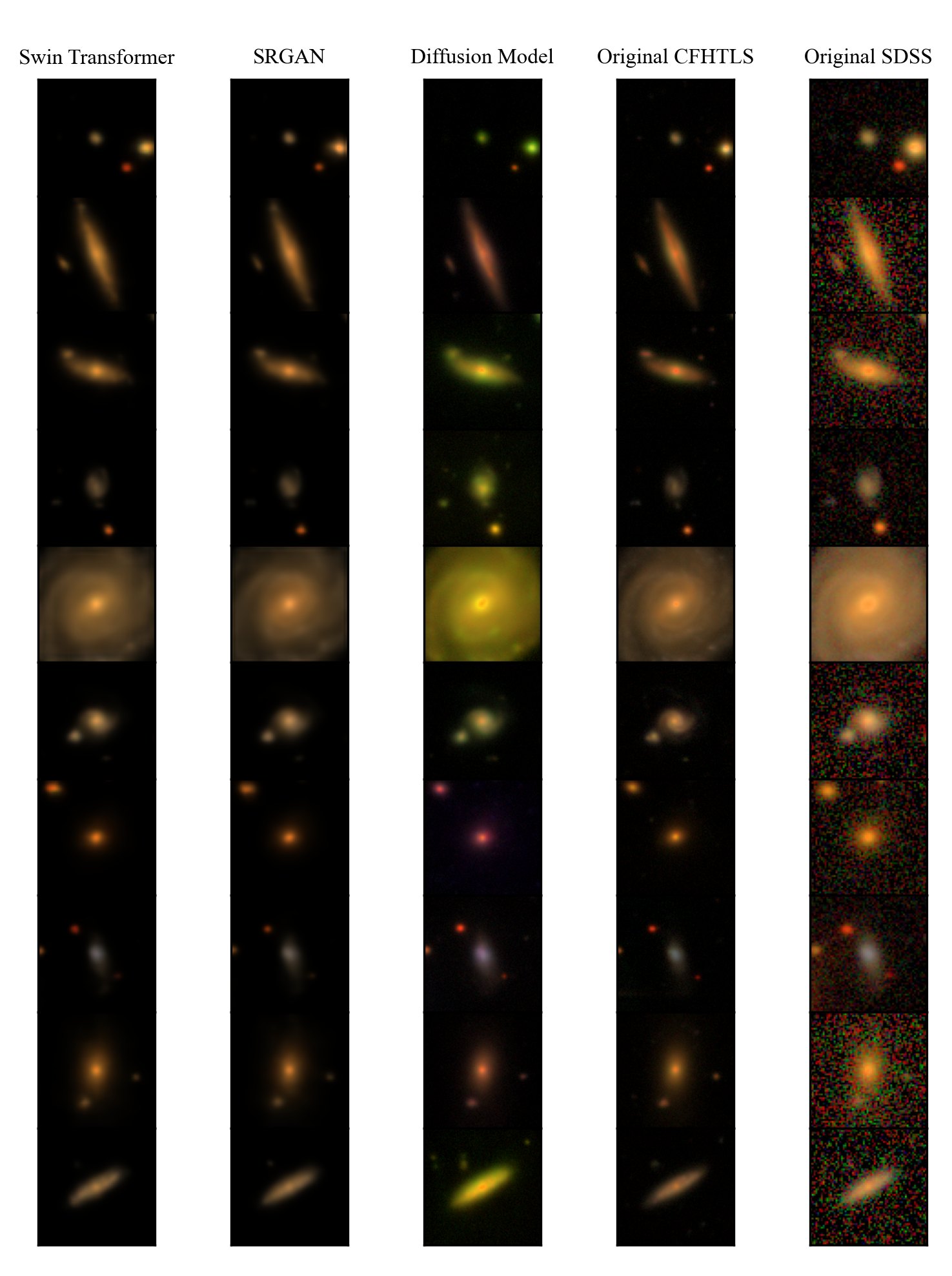}}
\caption{Exemplar RGB images generated by the S2C translation models (i.e. the Swin Transformer, the SRGAN, and the diffusion model) and the corresponding original CFHTLS validation images (target) and original SDSS validation images (input). The images were first restored to the initial flux scale using the reverse of Eq.~\ref{eq:norm} and then converted to RGB format using the method from \citet{Lupton2004}.}
\label{fig:rgb_s2c}
\end{center}
\end{figure*}

\end{CJK*}
\end{document}